\documentclass[preprint, amsmath, amssymb, aps,onecolumn] {revtex4-1}
\usepackage{graphicx}% Include figure files
\usepackage{dcolumn}% Align table columns on decimal point
\usepackage{bm}% bold math
\usepackage{caption}
\usepackage{color}
\newcommand{\BaCoGeO}{Ba$_2$CoGe$_2$O$_7$}
\newcommand{\BaMnGeO}{Ba$_2$MnGe$_2$O$_7$}

\begin{document}
%\preprint{APS/123-QED}
\title{Nontrivial temperature dependence of magnetic anisotropy in multiferroics Ba$_2$MnGe$_2$O$_7$}
\author{Shunsuke Hasegawa,$^1$ Shohei Hayashida,$^{1}$ Shinichiro Asai,$^1$ Masato Matsuura$^2$, Zaliznyak Igor$^3$ and Takatsugu Masuda$^{1,4,5}$ }
\affiliation{$^1$ Institute for Solid State Physics, The University of Tokyo, Chiba 277-8581, Japan \\
%$^2$ Laboratory for Solid State Physics, ETH Z\"{u}rich, CH-8093, Switzerland \\
$^2$ Neutron Science and Technology Center, Comprehensive Research Organization for Science and Society, Ibaraki 319-1106, Japan \\
$^3$ Condensed Matter Physics and Materials Science Department, Brookhaven National Laboratory, Upton, NY 11973, USA \\
$^4$ Institute of Materials Structure Science, High Energy Accelerator Research Organization, Ibaraki 305-0801, Japan \\
$^5$ Trans-scale Quantum Science Institute, The University of Tokyo, Tokyo 113-0033, Japan 
}
\date{\today}

\begin{abstract}
We measured the temperature dependences of the static magnetization and the spin excitation 
in a square-lattice multiferroics \BaMnGeO . %by combination of 
%the DC magnetometry and the inelastic neutron scattering techniques. 
An anisotropy gap of the observed low energy mode is scaled by electric polarization rather than a
power of sublattice moment. 
Spin nematic interaction in effective spin Hamiltonian, which is equivalent to interaction of electric polarization, 
is responsible for the easy-axis anisotropy. 
The nontrivial behavior of the anisotropy gap can be rationalized as change of the hybridized $d$-$p$ orbital 
with temperature, leading to the temperature dependence of the spin nematic interaction. 
\end{abstract}

\maketitle

Spin-driven multiferroics~\cite{RPP77.076501,ADVP64.519, NatRM116046F} have been extensively studied since 
the discovery of an enhanced magnetoelectric (ME) effect in TbMnO$_3$ \cite{Nature426.55}. 
Through spin-orbit coupling (SOC), a spin order induces a change of charge distribution, 
leading to emergence of electric polarization. 
The microscopic mechanisms of the multiferroics are categorized 
into three types~\cite{RPP77.076501}: the spin current \cite{PRL95.057205}, the exchange striction \cite{Cheong07} and the spin dependent $d$-$p$ hybridization \cite{JPSJ76.073702, PRB74.224444}.
A notable feature of the latter is that hybridized $d$ and $p$ orbitals of magnetic ion 
and ligand are modulated by spin states via SOC, and this induces an electric polarization. 
The relation between the electric polarization $\bm P$ and the spin moment $\bm S$ is 
locally described as $\bm{P} = \Lambda \sum_i ( \bm{S} \cdot \bm{e}_i ) ^2 \bm{e}_i$, 
where $\bm{e}_i$ is the bonding vector between the magnetic ion and the $i$-th ligand.
Coefficient $\Lambda$ is determined by SOC and transfer integrals between the magnetic ion and ligands.
The mechanism has been identified in CuFeO$_2$ \cite{JPSJ76.073702}, \aa kermanite compounds \cite{PRB85.174106, PRB84.224419, PRL112.127205}, Cu$_2$OSeO$_3$ \cite{Science336.198, PRL113.107203} and rare-earth 
ferroborates~\cite{PRB87.024413, PRB89.195126}.
Since the direction of the local spin moment determines the direction of the electric polarization, magnetic anisotropy plays a 
key role in forming multiferroic structure, $i.e.$, the simultaneous structures of the spin and polarization~\cite{PRB84.224419, PRL112.127205, PRB92.054402}.
The magnitude of the anisotropy gives the energy scale for the control of the magnetism by 
the electric field as well as magnetic field, which was tested in Cu$_2$OSeO$_3$~\cite{PRL113.107203} and 
an \aa kermanite Ba$_2$CoGe$_2$O$_7$~\cite{PRB94.094418}. 
Although the temperature ($T$) dependences of the anisotropy gap in magnon spectrum in a conventional magnet 
scales as 
a power of 
the sublattice magnetic moment~\cite{PR114.705, PRB.3.1736}, it may not be the case for the 
multiferroics. 
The change of the polarization with $T$ affects 
the spin-interaction parameters through the $d$-$p$ hybridization, which might lead to a nontrivial behavior in $T$ dependence 
of the anisotropy and a low-energy spin dynamics near the magnetic $\Gamma$ point. 

\begin{figure}[btph]
\includegraphics[width=8cm]{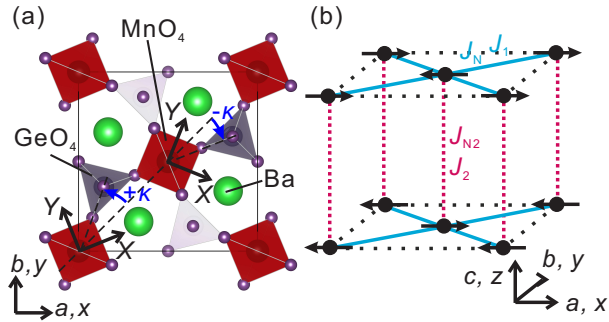}
\captionsetup{justification=raggedright}
\caption{
(a) Crystal structure of Ba$_2$MnGe$_2$O$_7$. MnO$_4$ tetrahedra are alternatively tilted by the characteristic angle of $\kappa$. 
(b) Magnetic structure of Ba$_2$MnGe$_2$O$_7$ and the exchange pathways. 
}
\label{fig0}
\end{figure}

We focus on a multiferroics Ba$_2$MnGe$_2$O$_7$~\cite{PRB81.100402} which is isostructural to \BaCoGeO\, 
but is distinct in that the effect of the crystal electric field on the isotropic charge distribution of the half-filled shell 
of Mn$^{2+}$ ion is small. 
Hence, weak spin-nematic interaction originating from the $d$-$p$ hybridization with the ligand, 
which can be sensitive to small change of the orbitals with the temperature, 
is the main source of magnetic anisotropy. 
%plays an important role in the low energy dynamics as well as 
%a bulk property at low temperatures. 
The crystal structure is tetragonal and the space group is $P\bar{4} 2_1m$~\cite{inorgchem.8b00058}, as 
shown in Fig.~\ref{fig0}(a). 
Mn$^{2+}$ ions carrying spin $S$ = 5/2 form a square lattice in the $ab$ plane. 
The magnetic susceptibility exhibits a typical behavior of a classical square-lattice Heisenberg
antiferromagnet with the interaction of 26 $\mu$eV. 
At the N\'eel temperature ($T_{\rm N}$) of 4 K, a magnetic long-range order 
with the propagation vector of $(1,~0,~1/2)$ sets in, and 
a collinear spin structure in which the spins lie in the $ab$ plane is realized as 
shown in Fig.~\ref{fig0}(b). 
The magnitude of the magnetic moment at 1.7 K is 4.66 $\mu _B$ and is close to the saturation moment. 
The electric polarization appears below $T_{\rm N}$ when the magnetic field is applied along 
the crystallographic [110] direction \cite{PRB85.174106} and the origin of the multiferroicity 
is explained by the spin dependent $d$-$p$ hybridization mechanism, which is also the main source of Mn$^{2+}$ spin anisotropy. 
Inelastic neutron scattering (INS) experiment at 1.7 K using conventional triple-axis spectrometer 
with the energy resolution of 0.1 meV determined the main parameters in 
the presumed isotropic spin Hamiltonian; 
the nearest-neighbor interaction in the $ab$ plane $J_1$ is 27.8 $\mu$eV, and 
the interplane interaction $J_2$ is 1.0 $\mu$eV. 
%ESR experiment reported an energy gap from a single-ion anisotropy of about 0.11 meV
%at 1.8 K~\cite{PRB.98.064416}. 
To reveals the magnetic anisotropy in the low-energy dynamics and 
its possible nontrivial behavior in $T$ dependence, 
we study the detailed field - temperature ($H$ - $T$) phase diagram and 
inelastic neutron scattering spectra using a state-of-art spectrometer with ultra-high resolution. 

\begin{figure}[tp]
\captionsetup{justification=raggedright}
\begin{center}
\includegraphics[width=8cm]{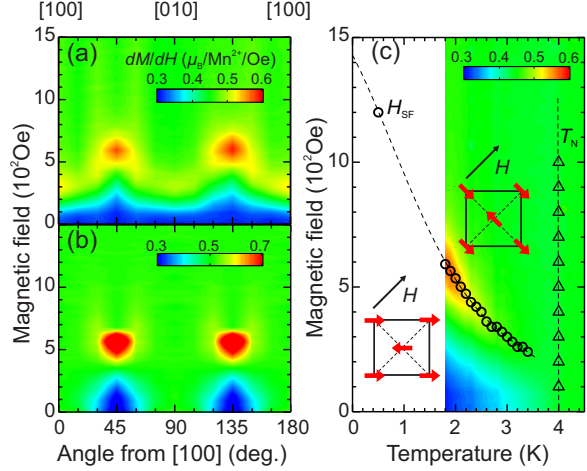}
\caption{Field derivative of magnetization ($dM/dH$). 
(a) False color plot of $dM/dH$ measured in the field applied in the $ab$ plane at 1.8 K. 
The horizontal axis is the angle between $[100]$ and the field directions. 
%$dM/dH$ curves in the field applied along [110] direction measured at typical temperatures are shown in 
%Supplementary Figs.~2(a), 2(c), and 2(e). 
(b) False color plot of calculated $dM/dH$ in the $ab$ plane. 
%The calculated $dM/dH$ curves are shown in Supplementary Figs.~2(b), 2(d), and 2(e). 
(c) Measured $H$ - $T$ map of $dM/dH$ for $H$ $\|$ [110]. 
$dM/dH$ curves measured at typical temperatures are shown in 
Supplementary Fig.~S3. 
Black circles indicate $H_{\rm SF}$ determined by the peak tops of $dM/dH$. 
A circle at 0.5 K indicates $H_{\rm SF}$ reported in previous study~\cite{PRB81.100402}. 
%A couple of blue circles are $H_{\rm SF}$ estimated by extrapolation and interpolation in Supplementary Fig.~S4. 
Triangles indicate N{\'e}el temperatures evaluated from the $T$ dependence of the magnetization. 
The representative descriptions of the magnetic structures below and above $H_{\rm SF}$ are inserted. }
\label{fig2}
\end{center}
\end{figure}

Single crystals were grown in O$_2$ atmosphere by the floating zone method.
The magnetization was measured by a commercial SQUID magnetometer.
The INS experiment was carried out using the 
near backscattering spectrometer DNA in J-PARC MLF \cite{JPSCP.8.036022}.
The horizontal scattering plane was the $ac$ plane, and the final energy of neutrons $E_{\rm f}$ was set to 2.084 meV by using Si (111) analyzer. 
The energy resolution was estimated to be 5.5 $\mu$eV in FWHM at an elastic condition of incoherent scattering 
near ${\bm Q} = (1, 0, 1/2)$. 
A dilution refrigerator was used to cool the sample to 0.05 K. 

In the field derivative of the magnetization $dM/dH$, the fourfold rotational symmetry in the 
$ab$ plane which is a characteristic feature of the spin-nematic interaction as demonstrated 
in \BaCoGeO ~\cite{PRL112.127205} is observed [Fig. \ref{fig2}(a)]. 
Enhancements of $dM/dH$ at $H$= 600 Oe for $H$ $\|$ [110] and [$\bar{1}$10] are due to 
a spin-flop (SF) transition.
The existence of the biaxial SF transitions suggests 
that the easy axes are [100] and [010]. 
The corresponding electric polarization structure is antiferroelectric where ${\bm P}$ is along the 
$c$ axis~\cite{PRB85.174106}. 
The SF field drastically decreases with the increasing $T$ as shown in $H$ - $T$ phase diagram in 
Fig.~\ref{fig2}(c), 
which contrasts with conventional antiferromagnets exhibiting approximately $T$ independent behavior 
at $T < T_{\rm N}$. 

\begin{figure*}[tp]
\captionsetup{justification=raggedright}
\includegraphics[width=17cm]{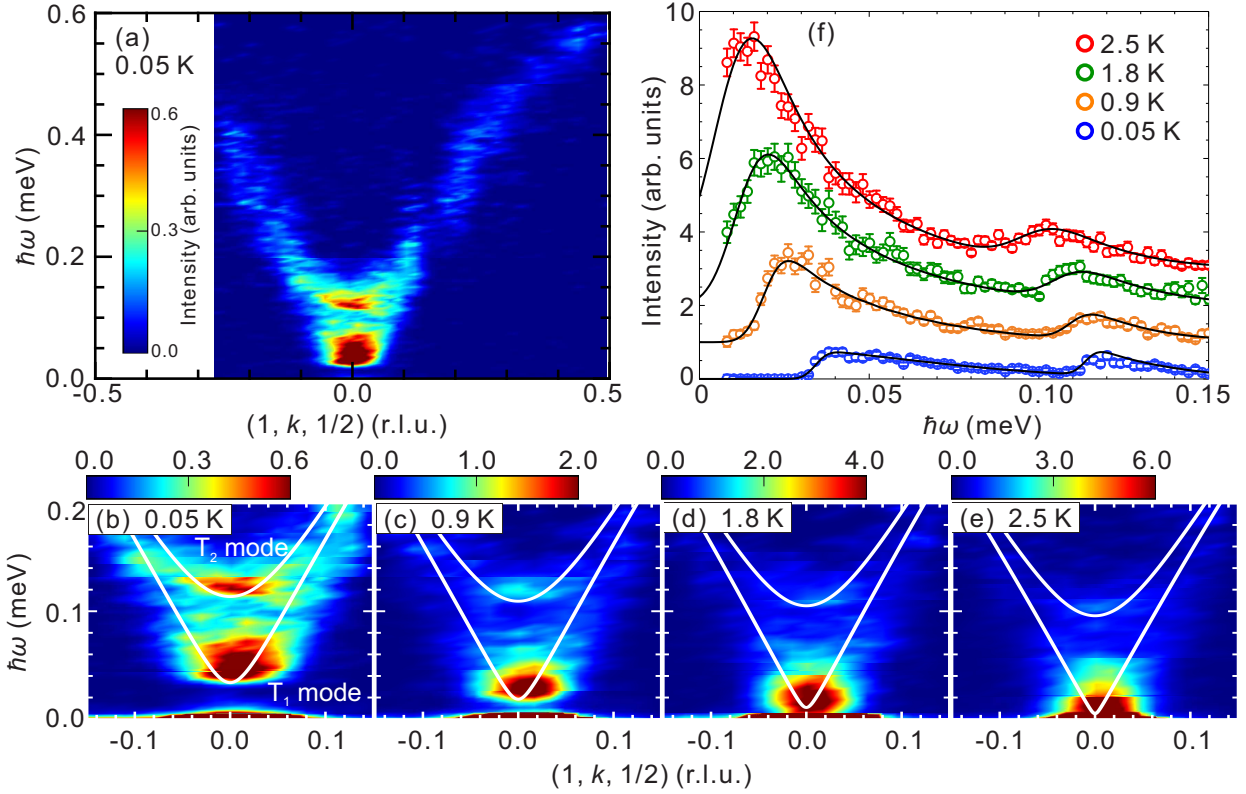}
\caption{Inelastic neutron scattering (INS) spectra. 
(a) False color plot of the INS spectrum measured at 0.05 K projected onto the $\hbar \omega$-$(1, k, 1/2)$ plane. The spectrum is integrated in the range of 0.9 $\leq h \leq$ 1.1 and 0.45 $\leq l \leq$ 0.55.
White solid curves are calculated magnon-dispersions. 
(b)-(e) The INS spectra focused on the low-energy range measured at (b) 0.05 K, (c) 0.9 K, (d) 1.8 K and (e) 2.5 K. 
White solid curves are calculated dispersions.
(f) Temperature evolution of constant-$\bm{q}$ cuts at $\bm{Q}=(1, 0, 1/2)$; data at different temperature are vertically offset. 
Coherent elastic scattering is subtracted for each temperature. 
The spectra are integrated in the range of -0.035 $\leq k \leq$ 0.035. 
%Vertical error bars indicate the statistical errors of 1 s.d. 
The asymmetric peak shape results from the wave vactor resolution. 
Black solid curves are the calculated magnon cross section. 
}
\label{fig1}
\end{figure*}

Figure \ref{fig1}(a) shows the INS spectrum measured at 0.05 K. 
Two dispersive modes with the boundary energy of 0.55 meV and with gaps $\sim 0.04$~meV and $\sim 0.1$~meV  
are clearly observed. 
The overall energy scale is consistent with the previous study~\cite{PRB81.100402}. 
It is remarkable that the two dispersive modes with finite gaps are clearly resolved, which was not the case in previous study 
because of the relaxed energy resolution.
Anisotropy gaps ($E_g$) of the low- and high-energy modes, T1 and T2, respectively, 
at the magnetic $\Gamma$ point, $\bm{Q}$ = (1, 0, 1/2), are identified at 
36 $\mu$eV and 113 $\mu$eV. 
The latter is consistent with the value reported in previous ESR study~\cite{PRB.98.064416}. 
$E_g$ of the ${\rm T}_1$ mode ($E_{g1}$) is drastically suppressed with the increase of $T$ as shown 
in Figs.~\ref{fig1}(b)-\ref{fig1}(e), while $E_g$ of the ${\rm T}_2$ mode ($E_{g2}$) is moderately suppressed. 
The intensities of the modes at the $\Gamma$ point increas with the increasing $T$ 
according to the thermal balance factor, as shown in Fig.~\ref{fig1}(f).

\begin{figure}[thb]
\captionsetup{justification=raggedright}
\includegraphics[width=8cm]{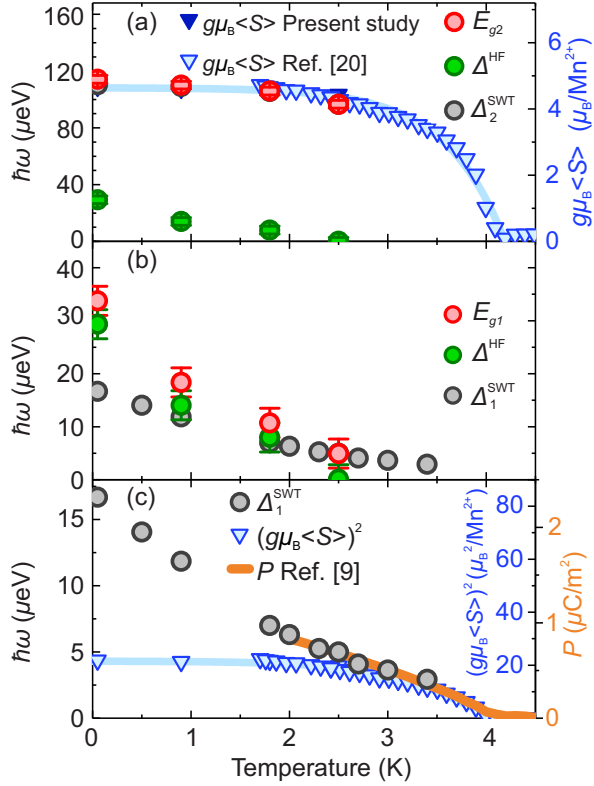}
\caption{The temperature dependences of gap energies observed in INS spectra, anisotropy energies, sublattice moment 
and electric polarization.
(a) $T$ dependences of $E_{g2}$ (red circles), $\Delta _2^{\rm SWT}$ (black circles), $\Delta^{\rm HF}$ (green circles) and sublattice moment (blue triangles from the present study and light blue ones from the previous 
study~\cite{PRB81.100402}). 
$E_{g2}$, $\Delta _2^{\rm SWT}$, and the sublattice moment are superimposed. 
The light blue curve is the guide to the eyes. 
%The sublattice moment in the present study 
%is estimated by scaling the integrated intensity of the magnetic Bragg peak of ${\bm Q}=(1, 0, 1/2)$ at 1.8 K 
%to the one in the previous study. 
(b) $T$ dependences of $E_{g1}$ (red circles), $\Delta _1^{\rm SWT}$ (black circles) and $\Delta^{\rm HF}$ (green circles). 
(c) $T$ dependences of $\Delta _1^{\rm SWT}$ (black circles), electric polarization $P$ (orange curve from Ref.~\cite{PRB85.174106}) and the second power of the sublattice moment (light blue triangles). 
The sublattice moment in this panel is a combination of 
measurements from the present and previous~\cite{PRB81.100402} studies in panel (a). 
Vertical error bars for $E_{g1}$, $E_{g2}$ and $\Delta^{\rm HF}$ in panels (a) and (b) 
indicate the experimental resolution of the spectrometer estimated at an elastic condition. 
}
\label{fig3}
\end{figure}

The $T$ dependences of $E_{g2}$ and the sublattice moment $g\mu _{\rm B}\langle S \rangle$ 
are shown in Fig.~\ref{fig3}(a). 
The change of $E_{g2}$ in the measurement range is small, and it scales as $g\mu _{\rm B}\langle S \rangle$. 
The behavior is consistent with conventional antiferromagnets
where single-ion anisotropy is dominated by quadratic forms of spin operators~\cite{PR114.705,PRB.3.1736}. 
In contrast, $E_{g1}$ in Fig.~\ref{fig3}(b) increases continuously with the decrease of 
$T$, and it cannot be scaled either by the sublattice moment or any power of it. 
This indicates that the T$_1$ mode is not purely magneticdynamics of electronic spin, but 
is hybridized with some other degrees of freedom. 

We calculate the $dM$/$dH$ curves and the dynamical structure factor of neutron scattering 
using the leading-order $1/S$ expansion of the following Hamiltonian:
\begin{eqnarray}
{\cal{H}} &=& \sum_{i,j} J_{1} \bm{S}_i \cdot \bm{S}_j + \sum_{k,l} J_{2} \bm{S}_k \cdot \bm{S}_l \nonumber \\
&&+ \sum_{i} \{ D (S_i^z)^2+g{\mu}_B\bm{S}_i \cdot \bm{H} \} + {\cal H}_N, \label{equation1} \\
{\cal H}_{\rm N} &=& \sum_{i,j} J_{\rm N} O_i^{XY} O_j^{XY}, \label{equation2}
\end{eqnarray}
where $O_i^{XY}=\cos (2 \kappa_i) (S_i^x S_i^y + S_i^y S_i^x ) - \sin(2 \kappa_i) \{ (S_i^x)^2-(S_i^y)^2 \}$. 
The $x,$ $y$ and $z$ axes in the Hamiltonian (\ref{equation1})
are along the crystallographic $a$-, $b$- and $c$-axes, respectively, as shown in Fig.~\ref{fig0}(b). 
$\kappa_i$ is a tilt angle of the MnO$_4$ tetrahedron from the crystallographic 
[110] direction as shown in Fig.~\ref{fig0}(a).
The first and the second terms are the nearest-neighbor intra- and interplane interactions of 
the square lattice, respectively. 
The third term is a single-ion anisotropy of an easy-plane type, with $D>0$, and the fourth is a Zeeman term. 
${\cal H}_{\rm N}$ is the spin-nematic interaction, 
which produces the biaxial anisotropy in the $ab$ plane~\cite{PRB84.224419, PRL112.127205}. 
Note that a single-ion anisotropy with the second order of spin operators is not allowed in 
the $ab$ plane due to the fourfold rotational symmetry. 
The lowest order in-plane single-ion anisotropy is of the fourth order and the reason it is not included is explained in the end of 
the supplementary sections I. 
Figure \ref{fig2}(b) shows the calculated angle dependence of $dM$/$dH$ using $J_{\rm N}$ = 0.12 neV and $g \mu _{\rm B} \langle S \rangle$ = 4.66 ${\mu}_{\rm B}$, which reasonably reproduces the experimental data at 1.8 K in Fig.~\ref{fig2}(a).
By using $H_{\rm SF}$ and $g \mu _{\rm B} \langle S \rangle$ measured at each temperature we can estimate 
an effective $J_{\rm N}(T)$. 
We find that $J_{\rm N}$ as well as $H_{\rm SF}$ is strongly dependent on $T$ [Supplementary Fig. S5] and 
it cannot be explained solely by the change of $g \mu _{\rm B} \langle S \rangle$. 
The details of the calculation are described in the supplementary sections I and II. 

Based on the spin-wave calculation, the gap energies of the magnetic anisotropy at the magnetic $\Gamma$-point are 
$\Delta_1^{\rm SWT}\sim 16\langle S(T) \rangle^2 \sqrt{J_{\rm N}(T)(2J_1+J_2 + D/2) }$ 
for the T$_1$ mode and 
$\Delta_2^{\rm SWT}=4\langle S(T) \rangle \sqrt{D(2J_1+J_2)}$ for the T$_2$ mode. 
$\Delta _1^{\rm SWT}$ can be calculated by using the values of $J_{\rm N}(T)$ from $dM/dH$ curve and $\langle S(T) \rangle$ from the neutron scattering experiment in the present study, and those of $J_1$, $J_2$ and $g$ in the previous study~\cite{PRB81.100402}. 
The obtained $T$ dependence of $\Delta _1^{\rm SWT}$ in Fig.~\ref{fig3}(b), however, is not
consistent with $E_{g1}$, particularly in the low $T$ region. 
Here we note that a Mn atom has a nuclear spin $I = 5/2$ and that the hyperfine coupling 
between nuclear and electron spins induces a low energy gap \cite{JETPL.1.134} in addition to 
the magnetic anisotropy gap. 
The energy gap induced by the hyperfine coupling has no $\bm{q}$-dependence. 
Then the magnon dispersion relation is modified; 
($\hbar \omega_i(\bm q))^2 = (\hbar \omega_i^{\rm{SWT}}(\bm q))^2 + (\Delta^{\rm{HF}})^2 $
where $\hbar \omega_i^{\rm{SWT}}(\bm q)$ is the pure magnon dispersion for T$_i$ mode ($i$ = 1,2) and $\Delta^{\rm{HF}}$ is the hyperfine gap \cite{JETPL.64.473}.
The two gaps $E_{g1}$ and $E_{g2}$ at $\bm{Q}=(1, 0, 1/2)$ are described as 
$E_{g1}^2 = \left( \Delta_1^{\rm SWT} \right)^2 + \left( \Delta^{\rm{HF}} \right)^2$ and 
$E_{g2}^2 = \left( \Delta_2^{\rm SWT} \right)^2 + \left( \Delta^{\rm{HF}} \right)^2, $
where 
$\Delta^{\rm{HF}}=\sqrt{ ( \chi_n(T) / \chi_c )} Ag {\mu}_B \langle S(T) \rangle$. 
Here $\chi_n(T)$ is the paramagnetic nuclear spin susceptibility following Curie law, $\chi_c$ is 
the electron spin susceptibility along the crystallographic $c$ direction, and $A$ is the hyperfine coupling. 
In order to estimate $\Delta ^{\rm HF}$ and $D$, we fit the constant-$\bm q$ cut 
in Fig.~\ref{fig1}(f) to the calculated magnon cross-section. 
The fit to the data is good as shown by black solid curves. 
The T$_1$ and T$_2$ modes 
in Figs.~\ref{fig1}(b)-\ref{fig1}(e) are reasonably reproduced by the calculated magnon-dispersions shown by the white curves. 
%The parameter $D$ is summarized in the supplementary Table SI. 
The obtained hyperfine coupling constant is $A$ = 240 kOe as described in the supplementary section IV, 
and it is consistent with Ref. \cite{JETPL.64.473} and the value reported in ${}^{55}$Mn$^{2+}$ ions in ZnF$_2$~\cite{Clogston}. 
The ${\rm T}_1$ mode is, thus, hybridized with the nuclear spin. 
%
%As described in the method section, both constant-$\bm q$ cuts in Fig.~\ref{fig1}f and 
%the INS spectra in Figs.~\ref{fig1}b-\ref{fig1}e are successfully analyzed by using the formulae; 
%$E_{gi}^2 = \left( \Delta_i^{\rm SWT} \right)^2 + \left( \Delta^{\rm{HF}} \right)^2$ ($i$ = 1,2)~\cite{JETPL.64.473}. 
%Here $\Delta^{\rm{HF}}$ is the induced gap by a hyperfine coupling. 
The $T$ dependence of $\Delta ^{\rm HF}$ is shown in Figs.~\ref{fig3}(a) and \ref{fig3}(b), 
and that of $\Delta _2^{\rm SWT}$ is shown in Fig.~\ref{fig3}(a). 
%In contrast with $\Delta_1^{\rm SWT}$, $\Delta _2^{\rm SWT}$ is quite close to $E_{g2}$, meaning that the 
%single-ion anisotropy $D$ dominates the energy gap. 
%Combination of the detailed analyses on $dM/dH$ curves and INS spectra leads to secure estimates of 
%$\Delta _1^{\rm SWT}$, $\Delta _2^{\rm SWT}$ and $\Delta^{\rm{HF}}$. 

$T$ dependences of $\Delta _1^{\rm SWT}$, a scaled electric polarization $P$~\cite{PRB85.174106} 
and the second power of the scaled sublattice moment $(g\mu _{\rm B}\langle S \rangle)^2$ are shown in Fig.~\ref{fig3}(c). 
We find that $\Delta _1^{\rm SWT}$ does not scale as $(g\mu _{\rm B}\langle S \rangle)^2$, 
but is rather consistent with the temperature dependence of electric polarization, $P$. 
To discuss the behavior, we rewrite the formula of ${\cal H}_{\rm N}$ in equation~(\ref{equation2}) 
in terms of $P$. 
From the local relation between $\bm S$ and $\bm P$, one has, $P_i^Z = \Lambda _ZO_i^{XY}$~\cite{PRB84.224419, PRL112.127205}. 
This relation leads to ${\cal H}_{\rm N} = \sum_{i,j} J_{\rm N} O_i^{XY} O_j^{XY} = \sum_{i,j} J_P P_i^Z P_j^Z$, where 
$J_P$ is an effective interaction between the electric polarizations; $J_P = J_{\rm N}/\Lambda _Z^2$.
%Then, we obtain $\Delta _1^{\rm SWT} = 16 \Lambda _Z\langle S(T) \rangle ^2 \sqrt{J_P(2J_1 + J_2 + D/2)}$. 
By considering the structures of the spin and electric polarization at $T < T_{\rm N}$, % structure is ${\bm S} = S(1, 0, 0)$ and the polarization structure is 
%${\bm P} = \sin (2\kappa ) \Lambda_Z S^2 (0, 0, 1)$ at $T < T_{\rm N}$, 
we obtain 
$\Delta _1^{\rm SWT} = 16 \langle P(T) \rangle \sqrt{J_P(2J_1 + J_2 + 1/2D)}/\sin (2\kappa)$. 
%The latter formula of $\Delta _1^{\rm SWT}$ and 
The fact that the magnetic anisotropy gap, $\Delta _1^{\rm SWT}(T)$, is scaled by the electric polarization, $P(T)$, means 
that $P$ induces an emergent magnetic field. 
In addition we notice that $J_P$ is $T$ independent. 
Drastic $T$ dependence of $J_{\rm N}$ thus results from the change of $\Lambda _Z$ with $T$; 
the modification of the hybridized orbitals of Mn$^{2+}$ and O$^{2-}$ ions with $T$ leads 
to a small change of the energy, which is negligible in highly anisotropic system but is enhanced in the nearly isotropic 
system such as the symmetric half-filled shell of Mn$^{2+}$ ion, and it is probed as the change of $J_{\rm N}$ thorough SOC. 
Since the change of the hybridized $d$-$p$ orbital with $T$ explicitly affects 
the anisotropy gap of the acoustic magnon, 
we call it {\it hybridized magnon} with the $d$-$p$ orbital. 
This is a characteristic quasiparticle of the multiferroics originating from the hybridization of spin and orbital momenta. 
%{\textcolor{red}{Since $H_{\rm SF}$ is equivalent to the anisotropy energy $\Delta _1^{\rm SWT}(T)$, $T$ dependence of $H_{\rm SF}$ is 
%a sensitive probe for SOC. 
%Reinvestigation of $H$ - $T$ phase diagrams in existing collinear antiferromagnets is 
%a new route to find magnetoelectric materials. }}

In contrast to $\Delta _1^{\rm SWT}$, the $T$ dependence of $\Delta_2^{\rm SWT}$ is almost flat at $T \le 2.5$ K, 
and the scaled $\Delta_2^{\rm SWT}$ is proportional to $g\mu _{\rm B}\langle S \rangle $ as shown in Fig.\ \ref{fig3}(a).
The result suggests that $D$ is essentially temperature independent. 
This is consistent with the dependence $\Delta_2^{\rm SWT}=4\langle S(T) \rangle \sqrt{D(2J_1+J_2)}$ describing 
conventional antiferromagnets in which a single-ion anisotropy is represented 
by the quadratic form of the spin operator~\cite{PR114.705, PRB.3.1736}. 

Now we compare the electric dipole energy of \BaMnGeO\ as a dielectric with that as a multiferroic. 
The former energy is estimated by the formula of the classical electromagnetism, 
$U_{(1,2)} = 1/(4\pi \epsilon _0 \epsilon r^3)({\bm P}_1 \cdot {\bm P}_2 - 3({\bm P}_1\cdot \hat{\bm r})({\bm P}_2\cdot \hat{\bm r}))$. 
The reported electric polarization is 
$(0, 0, p_c)$ with $p_c \sim 0.8~\mu {\rm C/m}^2$ at 2.0 K 
in the magnetic field of 1 T applied along the [110] direction~\cite{PRB85.174106}. 
The dielectric permittivity is $\epsilon \sim 14$~\cite{PRB.98.064416}.
We assume that the polarization is localized at each Mn$^{2+}$ ion.
By summing up the pairs of the nearest-neighbor Mn$^{2+}$ ions in the $ab$ plane and those along 
the $c$ axis, $U_{(1,2)}$ per Mn$^{2+}$ ion is estimated to be $-0.54$ peV. 
Next we estimate the electric dipole energy as a multiferroic system, 
where both dielectric and magnetic energies are renormalized. 
In \BaMnGeO\ the spin-nematic operator is
equivalent to the electric polarization, and only the component $P_i^Z = \Lambda _ZO_i^{XY}$ is active 
in the ordered state. 
Then the renormalized dipole energy is $E_{(1,2)} = J_{P}P_1^ZP_2^Z = J_{\rm N}O_1^{XY}O_2^{XY}$ at 1.8 K 
by using $J_{\rm N}$ and $g\mu _{\rm B}\langle S \rangle$ obtained in the present study. 
Note that the spin moment points along the $a$ or $b$ direction in the ordered state, $E_{(1,2)}$ per Mn$^{2+}$ ion 
is estimated to be $-7.95$ neV. 
The renormalized dipole energy in Ba$_2$MnGe$_2$O$_7$ is gigantic compared with the classical dipole energy. 
For the control of the electric polarization, the required electric field is much stronger than that estimated from 
the classical electromagnetism. 
The electric polarization in spin-driven multiferroics is, thus, robust to the external electric field.

In conclusion, we found that the temperature dependence of the magnetic anisotropy energy in the multiferroic \BaMnGeO\ 
is scaled by the electric polarization. 
The change of the hybridized orbitals of the metal and the ligands with the temperature 
determines the spin-nematic interaction, leading to nontrivial temperature dependence of 
the anisotropy energy in the neutron spectrum as well as that of the spin-flop field. 
The effect of hyperfine coupling between nuclear and electron spins is also observed in the low temperature region. 
Thorough account for this effect is important for accurate understanding of the temperature evolution of the electronic spin gap. 
%Reinvestigation of the spin-flop field in existing collinear antiferromagnets could be 
%a new route for search of magnetoelectric materials.

%{\bf Method}

%\begin{table}[p]
%\caption{\label{table1}Parameters obtained by linear spin wave theory in order to reproduce the spin wave including $E_g$.}
%\begin{ruledtabular}
%\begin{tabular}{cccccc}
%$T$ (K)& 0.05 & 0.5 & 0.9 & 1.8 & 2.5 \\ 
%\colrule
%$g\mu _{\rm B}\langle S \rangle$ ($\mu_B$) & 4.68 & -- & 4.62 & 4.66 & 4.43 \\
%$H_{\rm SF}$ (T) & 0.141 & -- & 0.101 & 0.059 & 0.040 \\
%$J_{\rm{N}}^{\rm{eff}}$ (neV) & 0.64 & -- & 0.34 & 0.12 & 0.07 \\
%$D$ ($\mu$eV) & 2.414 & -- & 2.424 & 2.230 & 2.065 \\
%$\Delta ^{\rm HF}$ ($\mu$eV) & 29.35 & -- & 14.05 & 7.99 & 0.083 \\
%$\Gamma$ ($\mu$eV) & 6.7 & -- & 10.3 & 15.5 & 18.8
%\label{tab2}
%\end{tabular}
%\end{ruledtabular}
%\end{table}

%\nocite{*}

%\bibliography{DNAver6}

\begin{thebibliography}{26}%
\makeatletter
\providecommand \@ifxundefined [1]{%
 \@ifx{#1\undefined}
}%
\providecommand \@ifnum [1]{%
 \ifnum #1\expandafter \@firstoftwo
 \else \expandafter \@secondoftwo
 \fi
}%
\providecommand \@ifx [1]{%
 \ifx #1\expandafter \@firstoftwo
 \else \expandafter \@secondoftwo
 \fi
}%
\providecommand \natexlab [1]{#1}%
\providecommand \enquote  [1]{``#1''}%
\providecommand \bibnamefont  [1]{#1}%
\providecommand \bibfnamefont [1]{#1}%
\providecommand \citenamefont [1]{#1}%
\providecommand \href@noop [0]{\@secondoftwo}%
\providecommand \href [0]{\begingroup \@sanitize@url \@href}%
\providecommand \@href[1]{\@@startlink{#1}\@@href}%
\providecommand \@@href[1]{\endgroup#1\@@endlink}%
\providecommand \@sanitize@url [0]{\catcode `\\12\catcode `\$12\catcode
  `\&12\catcode `\#12\catcode `\^12\catcode `\_12\catcode `\%12\relax}%
\providecommand \@@startlink[1]{}%
\providecommand \@@endlink[0]{}%
\providecommand \url  [0]{\begingroup\@sanitize@url \@url }%
\providecommand \@url [1]{\endgroup\@href {#1}{\urlprefix }}%
\providecommand \urlprefix  [0]{URL }%
\providecommand \Eprint [0]{\href }%
\providecommand \doibase [0]{http://dx.doi.org/}%
\providecommand \selectlanguage [0]{\@gobble}%
\providecommand \bibinfo  [0]{\@secondoftwo}%
\providecommand \bibfield  [0]{\@secondoftwo}%
\providecommand \translation [1]{[#1]}%
\providecommand \BibitemOpen [0]{}%
\providecommand \bibitemStop [0]{}%
\providecommand \bibitemNoStop [0]{.\EOS\space}%
\providecommand \EOS [0]{\spacefactor3000\relax}%
\providecommand \BibitemShut  [1]{\csname bibitem#1\endcsname}%
\let\auto@bib@innerbib\@empty
%</preamble>
\bibitem [{\citenamefont {Tokura}\ \emph {et~al.}(2014)\citenamefont {Tokura},
  \citenamefont {Seki},\ and\ \citenamefont {Nagaosa}}]{RPP77.076501}%
  \BibitemOpen
  \bibfield  {author} {\bibinfo {author} {\bibfnamefont {Y.}~\bibnamefont
  {Tokura}}, \bibinfo {author} {\bibfnamefont {S.}~\bibnamefont {Seki}}, \ and\
  \bibinfo {author} {\bibfnamefont {N.}~\bibnamefont {Nagaosa}},\ }\href
  {http://stacks.iop.org/0034-4885/77/i=7/a=076501} {\bibfield  {journal}
  {\bibinfo  {journal} {Rep. Prog. Phys.}\ }\textbf {\bibinfo {volume} {77}},\
  \bibinfo {pages} {076501} (\bibinfo {year} {2014})}\BibitemShut {NoStop}%
\bibitem [{\citenamefont {Dong}\ \emph {et~al.}(2015)\citenamefont {Dong},
  \citenamefont {Liu}, \citenamefont {Cheong},\ and\ \citenamefont
  {Ren}}]{ADVP64.519}%
  \BibitemOpen
  \bibfield  {author} {\bibinfo {author} {\bibfnamefont {S.}~\bibnamefont
  {Dong}}, \bibinfo {author} {\bibfnamefont {J.-M.}\ \bibnamefont {Liu}},
  \bibinfo {author} {\bibfnamefont {S.-W.}\ \bibnamefont {Cheong}}, \ and\
  \bibinfo {author} {\bibfnamefont {Z.}~\bibnamefont {Ren}},\ }\href {\doibase
  10.1080/00018732.2015.1114338} {\bibfield  {journal} {\bibinfo  {journal}
  {Adv. Phys.}\ }\textbf {\bibinfo {volume} {64}},\ \bibinfo {pages} {519}
  (\bibinfo {year} {2015})}\BibitemShut {NoStop}%
\bibitem [{\citenamefont {Fiebig}\ \emph {et~al.}(2016)\citenamefont {Fiebig},
  \citenamefont {Lottermoser}, \citenamefont {Meier},\ and\ \citenamefont
  {Trassin}}]{NatRM116046F}%
  \BibitemOpen
  \bibfield  {author} {\bibinfo {author} {\bibfnamefont {M.}~\bibnamefont
  {Fiebig}}, \bibinfo {author} {\bibfnamefont {T.}~\bibnamefont {Lottermoser}},
  \bibinfo {author} {\bibfnamefont {D.}~\bibnamefont {Meier}}, \ and\ \bibinfo
  {author} {\bibfnamefont {M.}~\bibnamefont {Trassin}},\ }\href {\doibase
  10.1038/natrevmats.2016.46} {\bibfield  {journal} {\bibinfo  {journal} {Nat.
  Rev. Mater.}\ }\textbf {\bibinfo {volume} {1}},\ \bibinfo {pages} {16046}
  (\bibinfo {year} {2016})}\BibitemShut {NoStop}%
\bibitem [{\citenamefont {Kimura}\ \emph {et~al.}(2003)\citenamefont {Kimura},
  \citenamefont {Goto}, \citenamefont {Shintani}, \citenamefont {Ishizaka},
  \citenamefont {hisa Arima},\ and\ \citenamefont {Tokura}}]{Nature426.55}%
  \BibitemOpen
  \bibfield  {author} {\bibinfo {author} {\bibfnamefont {T.}~\bibnamefont
  {Kimura}}, \bibinfo {author} {\bibfnamefont {T.~N.}\ \bibnamefont {Goto}},
  \bibinfo {author} {\bibfnamefont {H.}~\bibnamefont {Shintani}}, \bibinfo
  {author} {\bibfnamefont {K.}~\bibnamefont {Ishizaka}}, \bibinfo {author}
  {\bibfnamefont {T.}~\bibnamefont {hisa Arima}}, \ and\ \bibinfo {author}
  {\bibfnamefont {Y.}~\bibnamefont {Tokura}},\ }\href@noop {} {\bibfield
  {journal} {\bibinfo  {journal} {Nature}\ }\textbf {\bibinfo {volume} {426}},\
  \bibinfo {pages} {55} (\bibinfo {year} {2003})}\BibitemShut {NoStop}%
\bibitem [{\citenamefont {Katsura}\ \emph {et~al.}(2005)\citenamefont
  {Katsura}, \citenamefont {Nagaosa},\ and\ \citenamefont
  {Balatsky}}]{PRL95.057205}%
  \BibitemOpen
  \bibfield  {author} {\bibinfo {author} {\bibfnamefont {H.}~\bibnamefont
  {Katsura}}, \bibinfo {author} {\bibfnamefont {N.}~\bibnamefont {Nagaosa}}, \
  and\ \bibinfo {author} {\bibfnamefont {A.~V.}\ \bibnamefont {Balatsky}},\
  }\href {\doibase 10.1103/PhysRevLett.95.057205} {\bibfield  {journal}
  {\bibinfo  {journal} {Phys. Rev. Lett.}\ }\textbf {\bibinfo {volume} {95}},\
  \bibinfo {pages} {057205} (\bibinfo {year} {2005})}\BibitemShut {NoStop}%
\bibitem [{\citenamefont {Cheong}\ and\ \citenamefont
  {Mostovoy}(2007)}]{Cheong07}%
  \BibitemOpen
  \bibfield  {author} {\bibinfo {author} {\bibfnamefont {S.-W.}\ \bibnamefont
  {Cheong}}\ and\ \bibinfo {author} {\bibfnamefont {M.}~\bibnamefont
  {Mostovoy}},\ }\href@noop {} {\bibfield  {journal} {\bibinfo  {journal} {Nat.
  Mater.}\ }\textbf {\bibinfo {volume} {6}},\ \bibinfo {pages} {13} (\bibinfo
  {year} {2007})}\BibitemShut {NoStop}%
\bibitem [{\citenamefont {Arima}(2007)}]{JPSJ76.073702}%
  \BibitemOpen
  \bibfield  {author} {\bibinfo {author} {\bibfnamefont {T.-h.}\ \bibnamefont
  {Arima}},\ }\href {\doibase 10.1143/JPSJ.76.073702} {\bibfield  {journal}
  {\bibinfo  {journal} {Journal of the Physical Society of Japan}\ }\textbf
  {\bibinfo {volume} {76}},\ \bibinfo {pages} {073702} (\bibinfo {year}
  {2007})}\BibitemShut {NoStop}%
\bibitem [{\citenamefont {Jia}\ \emph {et~al.}(2006)\citenamefont {Jia},
  \citenamefont {Onoda}, \citenamefont {Nagaosa},\ and\ \citenamefont
  {Han}}]{PRB74.224444}%
  \BibitemOpen
  \bibfield  {author} {\bibinfo {author} {\bibfnamefont {C.}~\bibnamefont
  {Jia}}, \bibinfo {author} {\bibfnamefont {S.}~\bibnamefont {Onoda}}, \bibinfo
  {author} {\bibfnamefont {N.}~\bibnamefont {Nagaosa}}, \ and\ \bibinfo
  {author} {\bibfnamefont {J.~H.}\ \bibnamefont {Han}},\ }\href {\doibase
  10.1103/PhysRevB.74.224444} {\bibfield  {journal} {\bibinfo  {journal} {Phys.
  Rev. B}\ }\textbf {\bibinfo {volume} {74}},\ \bibinfo {pages} {224444}
  (\bibinfo {year} {2006})}\BibitemShut {NoStop}%
\bibitem [{\citenamefont {Murakawa}\ \emph {et~al.}(2012)\citenamefont
  {Murakawa}, \citenamefont {Onose}, \citenamefont {Miyahara}, \citenamefont
  {Furukawa},\ and\ \citenamefont {Tokura}}]{PRB85.174106}%
  \BibitemOpen
  \bibfield  {author} {\bibinfo {author} {\bibfnamefont {H.}~\bibnamefont
  {Murakawa}}, \bibinfo {author} {\bibfnamefont {Y.}~\bibnamefont {Onose}},
  \bibinfo {author} {\bibfnamefont {S.}~\bibnamefont {Miyahara}}, \bibinfo
  {author} {\bibfnamefont {N.}~\bibnamefont {Furukawa}}, \ and\ \bibinfo
  {author} {\bibfnamefont {Y.}~\bibnamefont {Tokura}},\ }\href {\doibase
  10.1103/PhysRevB.85.174106} {\bibfield  {journal} {\bibinfo  {journal} {Phys.
  Rev. B}\ }\textbf {\bibinfo {volume} {85}},\ \bibinfo {pages} {174106}
  (\bibinfo {year} {2012})}\BibitemShut {NoStop}%
\bibitem [{\citenamefont {Romh\'anyi}\ \emph {et~al.}(2011)\citenamefont
  {Romh\'anyi}, \citenamefont {Lajk\'o},\ and\ \citenamefont
  {Penc}}]{PRB84.224419}%
  \BibitemOpen
  \bibfield  {author} {\bibinfo {author} {\bibfnamefont {J.}~\bibnamefont
  {Romh\'anyi}}, \bibinfo {author} {\bibfnamefont {M.}~\bibnamefont {Lajk\'o}},
  \ and\ \bibinfo {author} {\bibfnamefont {K.}~\bibnamefont {Penc}},\ }\href
  {\doibase 10.1103/PhysRevB.84.224419} {\bibfield  {journal} {\bibinfo
  {journal} {Phys. Rev. B}\ }\textbf {\bibinfo {volume} {84}},\ \bibinfo
  {pages} {224419} (\bibinfo {year} {2011})}\BibitemShut {NoStop}%
\bibitem [{\citenamefont {Soda}\ \emph {et~al.}(2014)\citenamefont {Soda},
  \citenamefont {Matsumoto}, \citenamefont {M\aa{}nsson}, \citenamefont
  {Ohira-Kawamura}, \citenamefont {Nakajima}, \citenamefont {Shiina},\ and\
  \citenamefont {Masuda}}]{PRL112.127205}%
  \BibitemOpen
  \bibfield  {author} {\bibinfo {author} {\bibfnamefont {M.}~\bibnamefont
  {Soda}}, \bibinfo {author} {\bibfnamefont {M.}~\bibnamefont {Matsumoto}},
  \bibinfo {author} {\bibfnamefont {M.}~\bibnamefont {M\aa{}nsson}}, \bibinfo
  {author} {\bibfnamefont {S.}~\bibnamefont {Ohira-Kawamura}}, \bibinfo
  {author} {\bibfnamefont {K.}~\bibnamefont {Nakajima}}, \bibinfo {author}
  {\bibfnamefont {R.}~\bibnamefont {Shiina}}, \ and\ \bibinfo {author}
  {\bibfnamefont {T.}~\bibnamefont {Masuda}},\ }\href {\doibase
  10.1103/PhysRevLett.112.127205} {\bibfield  {journal} {\bibinfo  {journal}
  {Phys. Rev. Lett.}\ }\textbf {\bibinfo {volume} {112}},\ \bibinfo {pages}
  {127205} (\bibinfo {year} {2014})}\BibitemShut {NoStop}%
\bibitem [{\citenamefont {Seki}\ \emph {et~al.}(2012)\citenamefont {Seki},
  \citenamefont {Yu}, \citenamefont {Ishiwata},\ and\ \citenamefont
  {Tokura}}]{Science336.198}%
  \BibitemOpen
  \bibfield  {author} {\bibinfo {author} {\bibfnamefont {S.}~\bibnamefont
  {Seki}}, \bibinfo {author} {\bibfnamefont {X.~Z.}\ \bibnamefont {Yu}},
  \bibinfo {author} {\bibfnamefont {S.}~\bibnamefont {Ishiwata}}, \ and\
  \bibinfo {author} {\bibfnamefont {Y.}~\bibnamefont {Tokura}},\ }\href
  {\doibase 10.1126/science.1214143} {\bibfield  {journal} {\bibinfo  {journal}
  {Science}\ }\textbf {\bibinfo {volume} {336}},\ \bibinfo {pages} {198}
  (\bibinfo {year} {2012})}\BibitemShut {NoStop}%
\bibitem [{\citenamefont {White}\ \emph {et~al.}(2014)\citenamefont {White},
  \citenamefont {Pr\ifmmode~\check{s}\else \v{s}\fi{}a}, \citenamefont {Huang},
  \citenamefont {Omrani}, \citenamefont {\ifmmode \check{Z}\else
  \v{Z}\fi{}ivkovi\ifmmode~\acute{c}\else \'{c}\fi{}}, \citenamefont
  {Bartkowiak}, \citenamefont {Berger}, \citenamefont {Magrez}, \citenamefont
  {Gavilano}, \citenamefont {Nagy}, \citenamefont {Zang},\ and\ \citenamefont
  {R\o{}nnow}}]{PRL113.107203}%
  \BibitemOpen
  \bibfield  {author} {\bibinfo {author} {\bibfnamefont {J.~S.}\ \bibnamefont
  {White}}, \bibinfo {author} {\bibfnamefont {K.}~\bibnamefont
  {Pr\ifmmode~\check{s}\else \v{s}\fi{}a}}, \bibinfo {author} {\bibfnamefont
  {P.}~\bibnamefont {Huang}}, \bibinfo {author} {\bibfnamefont {A.~A.}\
  \bibnamefont {Omrani}}, \bibinfo {author} {\bibfnamefont {I.}~\bibnamefont
  {\ifmmode \check{Z}\else \v{Z}\fi{}ivkovi\ifmmode~\acute{c}\else
  \'{c}\fi{}}}, \bibinfo {author} {\bibfnamefont {M.}~\bibnamefont
  {Bartkowiak}}, \bibinfo {author} {\bibfnamefont {H.}~\bibnamefont {Berger}},
  \bibinfo {author} {\bibfnamefont {A.}~\bibnamefont {Magrez}}, \bibinfo
  {author} {\bibfnamefont {J.~L.}\ \bibnamefont {Gavilano}}, \bibinfo {author}
  {\bibfnamefont {G.}~\bibnamefont {Nagy}}, \bibinfo {author} {\bibfnamefont
  {J.}~\bibnamefont {Zang}}, \ and\ \bibinfo {author} {\bibfnamefont {H.~M.}\
  \bibnamefont {R\o{}nnow}},\ }\href {\doibase 10.1103/PhysRevLett.113.107203}
  {\bibfield  {journal} {\bibinfo  {journal} {Phys. Rev. Lett.}\ }\textbf
  {\bibinfo {volume} {113}},\ \bibinfo {pages} {107203} (\bibinfo {year}
  {2014})}\BibitemShut {NoStop}%
\bibitem [{\citenamefont {Popov}\ \emph {et~al.}(2013)\citenamefont {Popov},
  \citenamefont {Plokhov},\ and\ \citenamefont {Zvezdin}}]{PRB87.024413}%
  \BibitemOpen
  \bibfield  {author} {\bibinfo {author} {\bibfnamefont {A.~I.}\ \bibnamefont
  {Popov}}, \bibinfo {author} {\bibfnamefont {D.~I.}\ \bibnamefont {Plokhov}},
  \ and\ \bibinfo {author} {\bibfnamefont {A.~K.}\ \bibnamefont {Zvezdin}},\
  }\href {\doibase 10.1103/PhysRevB.87.024413} {\bibfield  {journal} {\bibinfo
  {journal} {Phys. Rev. B}\ }\textbf {\bibinfo {volume} {87}},\ \bibinfo
  {pages} {024413} (\bibinfo {year} {2013})}\BibitemShut {NoStop}%
\bibitem [{\citenamefont {Kurumaji}\ \emph {et~al.}(2014)\citenamefont
  {Kurumaji}, \citenamefont {Ohgushi},\ and\ \citenamefont
  {Tokura}}]{PRB89.195126}%
  \BibitemOpen
  \bibfield  {author} {\bibinfo {author} {\bibfnamefont {T.}~\bibnamefont
  {Kurumaji}}, \bibinfo {author} {\bibfnamefont {K.}~\bibnamefont {Ohgushi}}, \
  and\ \bibinfo {author} {\bibfnamefont {Y.}~\bibnamefont {Tokura}},\ }\href
  {\doibase 10.1103/PhysRevB.89.195126} {\bibfield  {journal} {\bibinfo
  {journal} {Phys. Rev. B}\ }\textbf {\bibinfo {volume} {89}},\ \bibinfo
  {pages} {195126} (\bibinfo {year} {2014})}\BibitemShut {NoStop}%
\bibitem [{\citenamefont {Hayashida}\ \emph {et~al.}(2015)\citenamefont
  {Hayashida}, \citenamefont {Soda}, \citenamefont {Itoh}, \citenamefont
  {Yokoo}, \citenamefont {Ohgushi}, \citenamefont {Kawana}, \citenamefont
  {R\o{}nnow},\ and\ \citenamefont {Masuda}}]{PRB92.054402}%
  \BibitemOpen
  \bibfield  {author} {\bibinfo {author} {\bibfnamefont {S.}~\bibnamefont
  {Hayashida}}, \bibinfo {author} {\bibfnamefont {M.}~\bibnamefont {Soda}},
  \bibinfo {author} {\bibfnamefont {S.}~\bibnamefont {Itoh}}, \bibinfo {author}
  {\bibfnamefont {T.}~\bibnamefont {Yokoo}}, \bibinfo {author} {\bibfnamefont
  {K.}~\bibnamefont {Ohgushi}}, \bibinfo {author} {\bibfnamefont
  {D.}~\bibnamefont {Kawana}}, \bibinfo {author} {\bibfnamefont {H.~M.}\
  \bibnamefont {R\o{}nnow}}, \ and\ \bibinfo {author} {\bibfnamefont
  {T.}~\bibnamefont {Masuda}},\ }\href {\doibase 10.1103/PhysRevB.92.054402}
  {\bibfield  {journal} {\bibinfo  {journal} {Phys. Rev. B}\ }\textbf {\bibinfo
  {volume} {92}},\ \bibinfo {pages} {054402} (\bibinfo {year}
  {2015})}\BibitemShut {NoStop}%
\bibitem [{\citenamefont {Soda}\ \emph {et~al.}(2016)\citenamefont {Soda},
  \citenamefont {Hayashida}, \citenamefont {Roessli}, \citenamefont
  {M\aa{}nsson}, \citenamefont {White}, \citenamefont {Matsumoto},
  \citenamefont {Shiina},\ and\ \citenamefont {Masuda}}]{PRB94.094418}%
  \BibitemOpen
  \bibfield  {author} {\bibinfo {author} {\bibfnamefont {M.}~\bibnamefont
  {Soda}}, \bibinfo {author} {\bibfnamefont {S.}~\bibnamefont {Hayashida}},
  \bibinfo {author} {\bibfnamefont {B.}~\bibnamefont {Roessli}}, \bibinfo
  {author} {\bibfnamefont {M.}~\bibnamefont {M\aa{}nsson}}, \bibinfo {author}
  {\bibfnamefont {J.~S.}\ \bibnamefont {White}}, \bibinfo {author}
  {\bibfnamefont {M.}~\bibnamefont {Matsumoto}}, \bibinfo {author}
  {\bibfnamefont {R.}~\bibnamefont {Shiina}}, \ and\ \bibinfo {author}
  {\bibfnamefont {T.}~\bibnamefont {Masuda}},\ }\href {\doibase
  10.1103/PhysRevB.94.094418} {\bibfield  {journal} {\bibinfo  {journal} {Phys.
  Rev. B}\ }\textbf {\bibinfo {volume} {94}},\ \bibinfo {pages} {094418}
  (\bibinfo {year} {2016})}\BibitemShut {NoStop}%
\bibitem [{\citenamefont {Johnson}\ and\ \citenamefont
  {Nethercot}(1959)}]{PR114.705}%
  \BibitemOpen
  \bibfield  {author} {\bibinfo {author} {\bibfnamefont {F.~M.}\ \bibnamefont
  {Johnson}}\ and\ \bibinfo {author} {\bibfnamefont {A.~H.}\ \bibnamefont
  {Nethercot}},\ }\href {\doibase 10.1103/PhysRev.114.705} {\bibfield
  {journal} {\bibinfo  {journal} {Phys. Rev.}\ }\textbf {\bibinfo {volume}
  {114}},\ \bibinfo {pages} {705} (\bibinfo {year} {1959})}\BibitemShut
  {NoStop}%
\bibitem [{\citenamefont {Birgeneau}\ \emph {et~al.}(1971)\citenamefont
  {Birgeneau}, \citenamefont {Skalyo},\ and\ \citenamefont
  {Shirane}}]{PRB.3.1736}%
  \BibitemOpen
  \bibfield  {author} {\bibinfo {author} {\bibfnamefont {R.~J.}\ \bibnamefont
  {Birgeneau}}, \bibinfo {author} {\bibfnamefont {J.}~\bibnamefont {Skalyo}}, \
  and\ \bibinfo {author} {\bibfnamefont {G.}~\bibnamefont {Shirane}},\ }\href
  {\doibase 10.1103/PhysRevB.3.1736} {\bibfield  {journal} {\bibinfo  {journal}
  {Phys. Rev. B}\ }\textbf {\bibinfo {volume} {3}},\ \bibinfo {pages} {1736}
  (\bibinfo {year} {1971})}\BibitemShut {NoStop}%
\bibitem [{\citenamefont {Masuda}\ \emph {et~al.}(2010)\citenamefont {Masuda},
  \citenamefont {Kitaoka}, \citenamefont {Takamizawa}, \citenamefont {Metoki},
  \citenamefont {Kaneko}, \citenamefont {Rule}, \citenamefont {Kiefer},
  \citenamefont {Manaka},\ and\ \citenamefont {Nojiri}}]{PRB81.100402}%
  \BibitemOpen
  \bibfield  {author} {\bibinfo {author} {\bibfnamefont {T.}~\bibnamefont
  {Masuda}}, \bibinfo {author} {\bibfnamefont {S.}~\bibnamefont {Kitaoka}},
  \bibinfo {author} {\bibfnamefont {S.}~\bibnamefont {Takamizawa}}, \bibinfo
  {author} {\bibfnamefont {N.}~\bibnamefont {Metoki}}, \bibinfo {author}
  {\bibfnamefont {K.}~\bibnamefont {Kaneko}}, \bibinfo {author} {\bibfnamefont
  {K.~C.}\ \bibnamefont {Rule}}, \bibinfo {author} {\bibfnamefont
  {K.}~\bibnamefont {Kiefer}}, \bibinfo {author} {\bibfnamefont
  {H.}~\bibnamefont {Manaka}}, \ and\ \bibinfo {author} {\bibfnamefont
  {H.}~\bibnamefont {Nojiri}},\ }\href {\doibase 10.1103/PhysRevB.81.100402}
  {\bibfield  {journal} {\bibinfo  {journal} {Phys. Rev. B}\ }\textbf {\bibinfo
  {volume} {81}},\ \bibinfo {pages} {100402(R)} (\bibinfo {year}
  {2010})}\BibitemShut {NoStop}%
\bibitem [{\citenamefont {Sazonov}\ \emph {et~al.}(2018)\citenamefont
  {Sazonov}, \citenamefont {Hutanu}, \citenamefont {Meven}, \citenamefont
  {Roth}, \citenamefont {Georgii}, \citenamefont {Masuda},\ and\ \citenamefont
  {N\'afr\'adi}}]{inorgchem.8b00058}%
  \BibitemOpen
  \bibfield  {author} {\bibinfo {author} {\bibfnamefont {A.}~\bibnamefont
  {Sazonov}}, \bibinfo {author} {\bibfnamefont {V.}~\bibnamefont {Hutanu}},
  \bibinfo {author} {\bibfnamefont {M.}~\bibnamefont {Meven}}, \bibinfo
  {author} {\bibfnamefont {G.}~\bibnamefont {Roth}}, \bibinfo {author}
  {\bibfnamefont {R.}~\bibnamefont {Georgii}}, \bibinfo {author} {\bibfnamefont
  {T.}~\bibnamefont {Masuda}}, \ and\ \bibinfo {author} {\bibfnamefont
  {B.}~\bibnamefont {N\'afr\'adi}},\ }\href {\doibase
  10.1021/acs.inorgchem.8b00058} {\bibfield  {journal} {\bibinfo  {journal}
  {Inorganic Chemistry}\ }\textbf {\bibinfo {volume} {57}},\ \bibinfo {pages}
  {5089} (\bibinfo {year} {2018})}\BibitemShut {NoStop}%
\bibitem [{\citenamefont {Shibata}\ \emph {et~al.}(2015)\citenamefont
  {Shibata}, \citenamefont {Takahashi}, \citenamefont {Kawakita}, \citenamefont
  {Matsuura}, \citenamefont {Yamada}, \citenamefont {Tominaga}, \citenamefont
  {Kambara}, \citenamefont {Kobayashi}, \citenamefont {Inamura}, \citenamefont
  {Nakatani}, \citenamefont {Nakajima},\ and\ \citenamefont
  {Arai}}]{JPSCP.8.036022}%
  \BibitemOpen
  \bibfield  {author} {\bibinfo {author} {\bibfnamefont {K.}~\bibnamefont
  {Shibata}}, \bibinfo {author} {\bibfnamefont {N.}~\bibnamefont {Takahashi}},
  \bibinfo {author} {\bibfnamefont {Y.}~\bibnamefont {Kawakita}}, \bibinfo
  {author} {\bibfnamefont {M.}~\bibnamefont {Matsuura}}, \bibinfo {author}
  {\bibfnamefont {T.}~\bibnamefont {Yamada}}, \bibinfo {author} {\bibfnamefont
  {T.}~\bibnamefont {Tominaga}}, \bibinfo {author} {\bibfnamefont
  {W.}~\bibnamefont {Kambara}}, \bibinfo {author} {\bibfnamefont
  {M.}~\bibnamefont {Kobayashi}}, \bibinfo {author} {\bibfnamefont
  {Y.}~\bibnamefont {Inamura}}, \bibinfo {author} {\bibfnamefont
  {T.}~\bibnamefont {Nakatani}}, \bibinfo {author} {\bibfnamefont
  {K.}~\bibnamefont {Nakajima}}, \ and\ \bibinfo {author} {\bibfnamefont
  {M.}~\bibnamefont {Arai}},\ }\href
  {https://journals.jps.jp/doi/abs/10.7566/JPSCP.8.036022} {\bibfield
  {journal} {\bibinfo  {journal} {JPS Conference Proceedings}\ }\textbf
  {\bibinfo {volume} {8}},\ \bibinfo {pages} {036022} (\bibinfo {year}
  {2015})}\BibitemShut {NoStop}%
\bibitem [{\citenamefont {Iguchi}\ \emph {et~al.}(2018)\citenamefont {Iguchi},
  \citenamefont {Nii}, \citenamefont {Kawano}, \citenamefont {Murakawa},
  \citenamefont {Hanasaki},\ and\ \citenamefont {Onose}}]{PRB.98.064416}%
  \BibitemOpen
  \bibfield  {author} {\bibinfo {author} {\bibfnamefont {Y.}~\bibnamefont
  {Iguchi}}, \bibinfo {author} {\bibfnamefont {Y.}~\bibnamefont {Nii}},
  \bibinfo {author} {\bibfnamefont {M.}~\bibnamefont {Kawano}}, \bibinfo
  {author} {\bibfnamefont {H.}~\bibnamefont {Murakawa}}, \bibinfo {author}
  {\bibfnamefont {N.}~\bibnamefont {Hanasaki}}, \ and\ \bibinfo {author}
  {\bibfnamefont {Y.}~\bibnamefont {Onose}},\ }\href {\doibase
  10.1103/PhysRevB.98.064416} {\bibfield  {journal} {\bibinfo  {journal} {Phys.
  Rev. B}\ }\textbf {\bibinfo {volume} {98}},\ \bibinfo {pages} {064416}
  (\bibinfo {year} {2018})}\BibitemShut {NoStop}%
\bibitem [{\citenamefont {Borovik-Romanov}\ and\ \citenamefont
  {Tulin}(1965)}]{JETPL.1.134}%
  \BibitemOpen
  \bibfield  {author} {\bibinfo {author} {\bibfnamefont {A.~S.}\ \bibnamefont
  {Borovik-Romanov}}\ and\ \bibinfo {author} {\bibfnamefont {V.}~\bibnamefont
  {Tulin}},\ }\href@noop {} {\bibfield  {journal} {\bibinfo  {journal} {Journal
  of Experimental and Theoretical Physics Lett.}\ }\textbf {\bibinfo {volume}
  {1}},\ \bibinfo {pages} {134} (\bibinfo {year} {1965})}\BibitemShut {NoStop}%
\bibitem [{\citenamefont {Zaliznyak}\ \emph {et~al.}(1996)\citenamefont
  {Zaliznyak}, \citenamefont {Zolin},\ and\ \citenamefont
  {Petrov}}]{JETPL.64.473}%
  \BibitemOpen
  \bibfield  {author} {\bibinfo {author} {\bibfnamefont {I.~A.}\ \bibnamefont
  {Zaliznyak}}, \bibinfo {author} {\bibfnamefont {N.~N.}\ \bibnamefont
  {Zolin}}, \ and\ \bibinfo {author} {\bibfnamefont {S.~V.}\ \bibnamefont
  {Petrov}},\ }\href@noop {} {\bibfield  {journal} {\bibinfo  {journal}
  {Journal of Experimental and Theoretical Physics Lett.}\ }\textbf {\bibinfo
  {volume} {64}},\ \bibinfo {pages} {473} (\bibinfo {year} {1996})}\BibitemShut
  {NoStop}%
\bibitem [{\citenamefont {Clogston}\ \emph {et~al.}(1960)\citenamefont
  {Clogston}, \citenamefont {Gordon}, \citenamefont {Jaccarino}, \citenamefont
  {Peter},\ and\ \citenamefont {Walker}}]{Clogston}%
  \BibitemOpen
  \bibfield  {author} {\bibinfo {author} {\bibfnamefont {A.~M.}\ \bibnamefont
  {Clogston}}, \bibinfo {author} {\bibfnamefont {J.~P.}\ \bibnamefont
  {Gordon}}, \bibinfo {author} {\bibfnamefont {V.}~\bibnamefont {Jaccarino}},
  \bibinfo {author} {\bibfnamefont {M.}~\bibnamefont {Peter}}, \ and\ \bibinfo
  {author} {\bibfnamefont {L.~R.}\ \bibnamefont {Walker}},\ }\href@noop {}
  {\bibfield  {journal} {\bibinfo  {journal} {Phys. Rev.}\ }\textbf {\bibinfo
  {volume} {117}},\ \bibinfo {pages} {1222} (\bibinfo {year}
  {1960})}\BibitemShut {NoStop}%
\end{thebibliography}

%

\clearpage

{\bf Acknowledgements}

We are grateful to Mr. T. Asami for supporting us in the neutron scattering experiment. 
The neutron experiment at the Materials and Life Science Experimental Facility of the J-PARC was performed under a user program (Proposal No. 2016B0140).
S. Hasegawa and S. Hayashida were supported by the Japan Society
for the Promotion of Science through the Leading Graduate Schools (MERIT). 
This project is supported by JSPS KAKENHI Grant Numbers 19KK0069 and 20K20896, 
and by US-Japan collaboration program.

\end{document}

%\preprint{APS/123-QED}
\title{Supplementary materials for Nontrivial temperature dependence of magnetic anisotropy in multiferroics Ba$_2$MnGe$_2$O$_7$ }
\author{Shunsuke Hasegawa,$^1$ Shohei Hayashida,$^{1}$ Shinichiro Asai,$^1$ Masato Matsuura$^2$, Zaliznyak Igor$^3$ and Takatsugu Masuda$^{1,4,5}$ }
\affiliation{$^1$ Institute for Solid State Physics, The University of Tokyo, Chiba 277-8581, Japan \\
%$^2$ Laboratory for Solid State Physics, ETH Z\"{u}rich, CH-8093, Switzerland \\
$^2$ Neutron Science and Technology Center, Comprehensive Research Organization for Science and Society, Ibaraki 319-1106, Japan \\
$^3$ Condensed Matter Physics and Materials Science Department, Brookhaven National Laboratory, Upton, NY 11973, USA \\
$^4$ Institute of Materials Structure Science, High Energy Accelerator Research Organization, Ibaraki 305-0801, Japan \\
$^5$ Trans-scale Quantum Science Institute, The University of Tokyo, Tokyo 113-0033, Japan 
}
\date{\today}

%\begin{abstract}
%Calculation of dynamical structure factor of \BNGO\ by linear spin-wave theory is summarized. 
%\end{abstract}

%\pacs{Valid PACS appear here}

\maketitle

%\section{Crystal structure of \BNGO\ }
%\begin{figure}[b]
%\includegraphics[width=8cm]{Sfig0.eps}
%\captionsetup{labelformat=empty,labelsep=none, justification=raggedright}
%\caption{{\bf Fig. S1}
%Crystal structure of \BNGO . 
%}
%\label{fig0}
%\end{figure}
%The crystal structure of \BNGO\ is shown in Fig. S\ref{fig0}. The crystal system is tetragonal with the space group $P\bar{4} 2_1m$~\cite{inorgchem.8b00058}. 
%Mn$^{2+}$ ions carrying spin $S$ = 5/2 form a square lattice in the crystallographic $ab$-plane. 
%MnO$_4$ tetrahedra are alternatively tilted by the characteristic angle of $\kappa$. 
%The coordinate defined by the capital $X$ and $Y$ is used for the Stevens operator 
%in Eq. (1) in the main article.   

\section{Spin-wave calculation for inelastic neutron scattering cross section of \BNGO\ }
\begin{figure}[b]
\includegraphics[width=8cm]{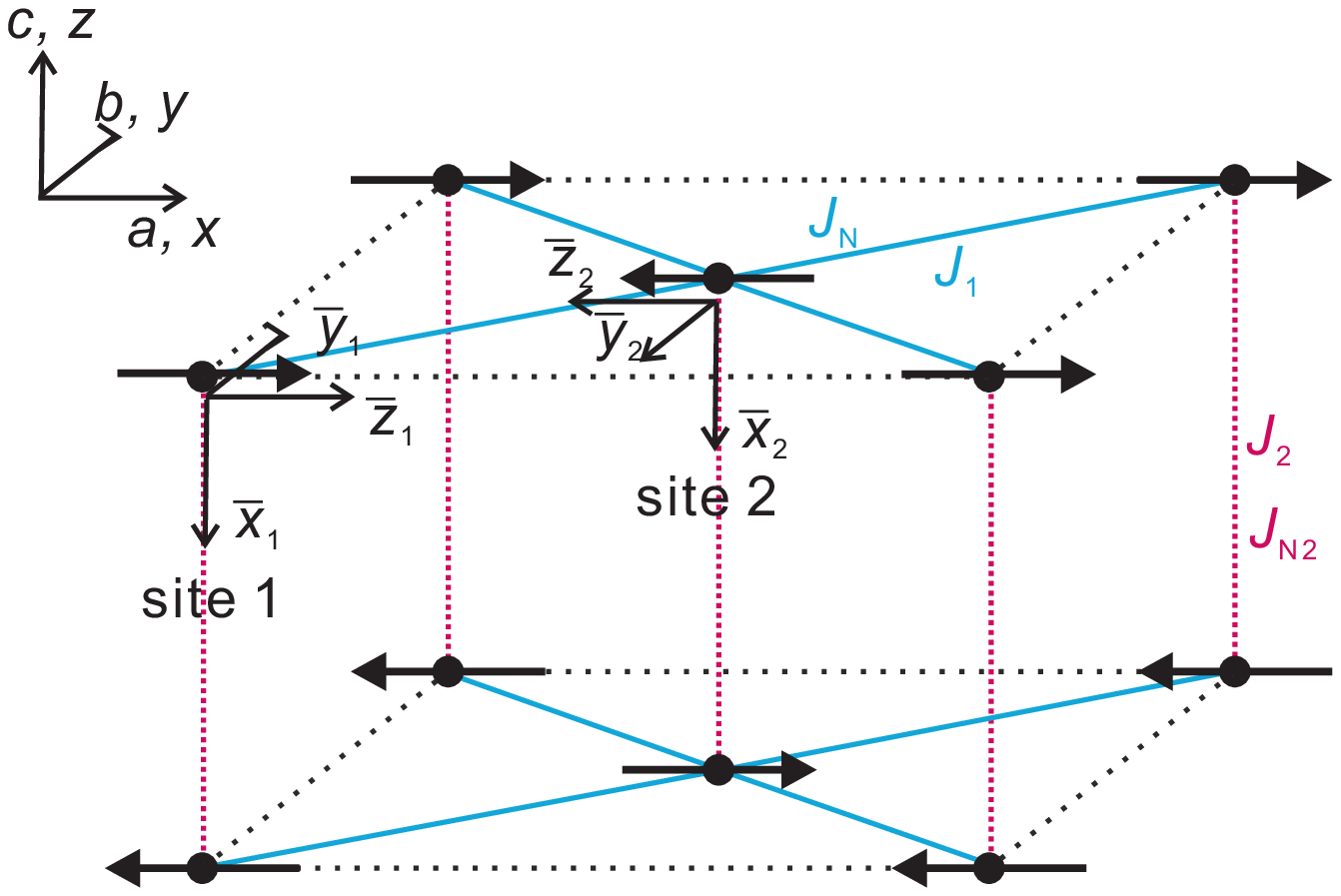}
\captionsetup{labelformat=empty,labelsep=none, justification=raggedright}
\caption{{\bf Fig. S1}
Magnetic structure of \BNGO\ and relation between global coordinate $x,y,z$ and local coordinate ${\bar x_i},{\bar y_i},{\bar z_i}$ ($i$ = 1, 2). 
}
\label{fig1}
\end{figure}

The dynamical structure factor of \BNGO\ is calculated by linear spin-wave theory using the recipe described in Ref.~\onlinecite{Fishman}. 
We start from the following spin Hamiltonian: 
\begin{eqnarray}
\label{equation2}
{\cal H} &=& {\cal H}_1 + {\cal H}_2 + {\cal H}_3 + {\cal H}_4, \label{hamiltonian} \\
{\cal H}_1 &=& \sum _{i,j}^{ab} J_1{\bm S}({\bm R}_i^{(1)}) \cdot {\bm S}({\bm R}_j^{(2)}) + \sum _{i,j}^{c} J_2{\bm S}({\bm R}_i^{(1)}) \cdot {\bm S}({\bm R}_j^{(2)}), \label{H1} \\
{\cal H}_2 &=& D \sum _i (S^z({\bm R}_i^{(r)}))^2, \\
{\cal H}_3 &=& J_{\rm N} \sum_{i,j}^{ab} O_{XY}({\bm R}_i^{(1)})O_{XY}({\bm R}_j^{(2)}) + J_{\rm N2} \sum_{i,j}^{c}O_{XY}({\bm R}_i^{(1)})O_{XY}({\bm R}_j^{(2)}), \label{H3} \\ \nonumber
&& O_{XY}({\bm R}_i^{(r)}) = \cos (2\kappa_i^{(r)})O_{xy}({\bm R}_i^{(r)}) - \sin (2\kappa_i^{(r)})O_2^2({\bm R}_i^{(r)}), \\ \nonumber
&& O_{xy}({\bm R}_i^{(r)}) = S^x({\bm R}_i^{(r)})S^y({\bm R}_i^{(r)}) + S^y({\bm R}_i^{(r)})S^x({\bm R}_i^{(r)}), \\ \nonumber
&& O_2^2 ({\bm R}_i^{(r)}) = (S^x({\bm R}_i^{(r)}))^2 - (S^y({\bm R}_i^{(r)}))^2, \label{H2} \\
{\cal H}_4 & = & \sum_i a\left( (S^x({\bm R}_i^{(r)}))^4 + (S^y({\bm R}_i^{(r)}))^4 \right) + b(S^x({\bm R}_i^{(r)}))^2(S^y({\bm R}_i^{(r)}))^2. \label{H4} 
\end{eqnarray}
Here ${\bm R}_i^{(r)}$ indicates a position vector of a spin with a site $r$ in a unit cell $i$. 
$\kappa_i^{(r)}$ is a tilt angle of MnO$_4$ tetrahedron from the crystallographic [110] direction 
as shown in Fig. 1(a) in the main article. 
Exchange pathways in ${\cal H}_1$ are shown in Fig.~S\ref{fig1}, and both of inplane interaction $J_1$ and interplane interaction 
$J_2$ are antiferromagnetic; $J_{1},J_2 > 0$. 
The single-ion anisotropy of the 2nd order in ${\cal H}_2$ is easy-plane type; $D > 0$. 
The inplane spin-nematic interaction $J_{\rm N}$ in ${\cal H}_3$ is antiferronematic, and 
the interplane spin-nematic interaction $J_{\rm N2}$ is ferronematic; $J_{\rm N} > 0$ and $J_{\rm N2} <0$. 
The anisotropy of the 4th order in ${\cal H}_4$ is biaxial anisotropy along $x$ and $y$ directions; $a<0$ and $b>0$. 
The $x$, $y$, and $z$ axes in the global coordinate correspond to the crystallographic $a$-, $b$-, and $c$-axes. 
The relation between the local coordinate at each sublattice (${\bar x}$, ${\bar y}$, ${\bar z}$) and the global coordinate 
$(x,~y,~z)$ is shown in Fig.~S\ref{fig1}. 
The spin operator in the global coordinate ${\bm S}({\bm R}_i^{(r)})$ and the one in the local coordinate ${\bar {\bm S}({\bm R}_i^{(r)})}$ is related 
by the unitary matrix $U_r$ as follows: 
\begin{eqnarray}
{\bm S}({\bm R}_i^{(r)}) &=& U_r^{-1} {\bar {\bm S}({\bm R}_i^{(r)})}, \\
U_1^{-1} &=& \left(
\begin{array}{ccc}
0 & 0 & 1 \\
0 & 1 & 0 \\
-1 & 0 & 0
\end{array}
\right), \\
U_2^{-1} &=& \left(
\begin{array}{ccc}
0 & 0 & -1 \\
0 & -1 & 0 \\
-1 & 0 & 0
\end{array}
\right). 
\end{eqnarray}
Holstein-Primakoff transformation yields the relation between local spin and boson operators: 
\begin{eqnarray}
{\bar S^x({\bm R}_i^{(r)})} &=& \sqrt{S/2}(a_i^{(r)}+a_i^{(r)\dagger}), \\
{\bar S^y({\bm R}_i^{(r)})} &=& -i\sqrt{S/2}(a_i^{(r)}+a_i^{(r)\dagger}), \\
{\bar S^z({\bm R}_i^{(r)})} &=& S-a_i^{(r)}a_i^{(r)\dagger}.
\end{eqnarray}
Fourier transformation of the boson operators are given by
\begin{eqnarray}
a_{\bm q}^{(r)} &=& 1/\sqrt{N_u}\sum_i^{(r)} e^{-i{\bm q}\cdot {\bm R}_i^{(r)}}a_i^{(r)}, \\
a_{\bm q}^{(r)\dagger} &=& 1/\sqrt{N_u}\sum_i^{(r)} e^{i{\bm q}\cdot {\bm R}_i^{(r)}}a_i^{(r)\dagger}, \\
a_i^{(r)} &=& 1/\sqrt{N_u}\sum_{\bm q}' e^{i{\bm q}\cdot {\bm R}_i^{(r)}}a_{\bm q}^{(r)}, \label{air}\\
a_i^{(r)\dagger} &=& 1/\sqrt{N_u}\sum_{\bm q}' e^{-{i{\bm q}\cdot {\bm R}_i^{(r)}}}a_{\bm q}^{(r)\dagger}. \label{aird}
\end{eqnarray}
The symbol $'$ in Eq.~(\ref{air}) and (\ref{aird}) means that the sums are taken over the 1st Brillouin zone. 
We transform the Hamiltonian in Eq.~(\ref{hamiltonian}) into the boson representation, and the second-order terms are 
obtained straightforwardly: 
\begin{eqnarray}
{\cal H} &=& \sum_{\bm q} '{\bm v}_{\bm q}^{\dagger} L {\bm v}_{\bm q}, \label{Hamiltonian2}\\
{\bm v}_{\bm q} &=& (a_{\bm q}^{(1)}, a_{\bm q}^{(2)}, a_{\bm -q}^{(1)\dagger}, a_{\bm -q}^{(2)\dagger}), \\
L & = & \left(
\begin{array}{cccc}
A & Q_{\bm q} & P & B_{\bm q} \\
Q_{\bm q} & A & B_{\bm q} & P \\
P & B_{\bm q} & A & Q_{\bm q} \\
B_{\bm q} & P & Q_{\bm q} & A
\end{array}
\right), \\
A&=& 2(2J_1+J_2)S+DS/2 - (-3K + 3a/2  + b/2 )S^3, \\
B_{\bm q} &=& 2(2J_1 \Gamma _{\bm q} ^{ab} + J_2 \Gamma _{\bm q}^c)S - Q_{\bm q}, \\
P&=&DS/2 + (-K-b/2)S^3, \\
K &=& 2(2J_{\rm N} - J_{\rm N2})\sin^2(2\kappa), \\
Q_{\bm q} &=& 2\cos^2(2\kappa)(4J_{\rm N} \Gamma _{\bm q}^{ab}+2J_{\rm N2}\Gamma _{\bm q}^c)S^3, \\
\Gamma _{\bm q}^{ab} &=& 1/2 \left( \cos \frac{a}{2}(q_x+q_y)+\cos \frac{a}{2} (q_x-q_y) \right), \\
\Gamma _{\bm q}^c &=& \cos(cq_z). 
\end{eqnarray} 
From the commutation rule of boson operators $a_{\bm q}^{(r)}$ and $a_{\bm q'}^{(s)\dagger}$, it is found that 
\begin{eqnarray}
[{\bm v_q}, {\bm v_{q'}}^{\dagger}] &=& \underline{N} \delta_{\bm q, q'}, \\
\underline{N}&=& \left(
\begin{array}{cc}
\underline{I} & 0 \\
0 & -\underline{I}
\end{array}
\right). 
\end{eqnarray} 
Here $\underline{I}$ is 2-dimensional unit matrix. 
A new matrix ${\cal L}$ is defined 
\begin{equation}
{\cal L} = L\cdot \underline{N}, 
\end{equation} 
and the dispersion relations of the one-magnon is obtained by the characteristic equation, 
\begin{equation}
\sum _j \left( {\cal L}_{n,j}(\bm q) - \delta _{n,j} \epsilon _n (\bm q) \right) X_{n,j}({\bm q})^* = 0, 
\end{equation}
where $\hbar \omega ^{\rm SWT}_n ({\bm q}) = 2 \epsilon _n(\bm q)$. 
The solutions are
\begin{eqnarray}
\hbar \omega ^{\rm SWT}_{1,2} (\bm q) &=& 2\sqrt{(A\mp Q)^2-(B\mp P)^2}, \label{creation}\\
\hbar \omega ^{\rm SWT}_{3,4} (\bm q) &=& -2\sqrt{(A\pm Q)^2-(B\pm P)^2}. \label{annihilation} 
\end{eqnarray}
Eqs.~(\ref{creation}) and (\ref{annihilation}) correspond to creation and annihilation of magnons, respectively. 
$\hbar \omega^{\rm SWT}_1(\bm q)$ is low energy mode, and $\hbar \omega^{\rm SWT}_2(\bm q)$ is high energy mode. 
The eigenvectors give the matrix $X$, and its inversion is as follows
\begin{equation}
X^{-1} = \left(
\begin{array}{cccc}
\frac{B-P}{4c_1\epsilon_1}, & \frac{B+P}{4c_2\epsilon_2}, & \frac{-(B+P)}{4c_3\epsilon_2}, &\frac{-(B-P)}{4c_4\epsilon_1} \\
\frac{-(B-P)}{4c_1\epsilon_1}, & \frac{B+P}{4c_2\epsilon_2}, & \frac{-(B+P)}{4c_3\epsilon_2}, &\frac{B-P}{4c_4\epsilon_1} \\
\frac{A-Q-\epsilon_1}{4c_1\epsilon_1}, & \frac{-A-Q+\epsilon_2}{4c_2\epsilon_2}, & \frac{A+Q+\epsilon_2)}{4c_3\epsilon_2}, &\frac{-A+Q-\epsilon_1}{4c_4\epsilon_1} \\
\frac{-A+Q+\epsilon_1}{4c_1\epsilon_1}, & \frac{-A-Q+\epsilon_2}{4c_2\epsilon_2}, & \frac{A+Q+\epsilon_2)}{4c_3\epsilon_2}, &\frac{A-Q+\epsilon_1}{4c_4\epsilon_1}
\end{array}
\right). 
\label{X-1}
\end{equation}
Normalization condition required from the commutation rule of the boson is $X\underline{N}X^{\dagger} = {\underline N},$ and 
it leads to 
\begin{eqnarray}
|c_1|^2 &=& \frac{1}{2\left( \left(\frac{A-Q+\epsilon_1}{B-P}\right)^2-1\right)} \label{c1}, \\
|c_2|^2 &=& \frac{1}{2\left( \left(\frac{A+Q+\epsilon_2}{B+P}\right)^2-1\right)} \label{c2}, \\
|c_3|^2 &=& \frac{1}{2\left( \left(\frac{A+Q-\epsilon_2}{B+P}\right)^2-1\right)} \label{c3}, \\
|c_4|^2 &=& \frac{1}{2\left( \left(\frac{A-Q-\epsilon_1}{B-P}\right)^2-1\right)} \label{c4}.
\end{eqnarray}

%The dynamical structure factor is obtained 
%\begin{eqnarray}
%S^{\alpha \beta}({\bm q}, \omega) &=& \sum_n S^{(n),\alpha \beta}(\bm q)\delta(\hbar \omega - \hbar \omega _{n}(\bm q)) \\
%S^{(n)\alpha \beta}(\bm q) &=& S/4 \sum_{r,s=1}^{2}W_{r,\alpha}^{(n)}(\bm q)W_{s,\beta}^{(n)}(\bm q)^* \\
%W_{r,\alpha}^{(n)}(\bm q) &=& V_{r, \alpha}^- X_{r,n}^{-1}+V_{r, \alpha}^+X_{r+2,n}^{-1}(\bm q) \\
%V_{r,\alpha}^{\pm}&=&(U_r^{-1})_{\alpha, x} \pm i(U_r)^{-1}_{\alpha, y}. 
%\end{eqnarray}

As is explained in the main article, in the case of a Mn$^{2+}$ ion having a nuclear spin $I$ = 5/2, hyperfine coupling 
between the nuclear and electron 
spins induces a low energy gap in addition to the magnetic anisotropy gap. 
The excitation energy modified by the hyperfine coupling is: 
\begin{equation}
\hbar \omega _n (\bm q)^2 = (\hbar \omega ^{\rm SWT}_n(\bm q))^2 + (\Delta^{\rm{HF}})^2, 
\label{DeltaHF}
\end{equation}
where $\hbar \omega _n (\bm q)$ is the magnon dispersion of \BNGO\ 
and $\Delta^{\rm{HF}}$ is the hyperfine gap \cite{JETPL.64.473, JETP.85.1035}. 
$\epsilon _{n}$ in Eqs.~(\ref{X-1})-(\ref{c4}) is replaced by $\epsilon '_{n}$, where 
\begin{equation}
\epsilon '_{n}=1/2\hbar \omega _n (\bm q). 
\end{equation}
Then the matrix $X$ becomes 
\begin{equation}
X'^{-1} = \left(
\begin{array}{cccc}
\frac{B-P}{4c '_1\epsilon '_1}, & \frac{B+P}{4c '_2\epsilon '_2}, & \frac{-(B+P)}{4c '_3\epsilon '_2}, &\frac{-(B-P)}{4c '_4\epsilon '_1} \\
\frac{-(B-P)}{4c '_1\epsilon '_1}, & \frac{B+P}{4c '_2\epsilon '_2}, & \frac{-(B+P)}{4c '_3\epsilon '_2}, &\frac{B-P}{4c '_4\epsilon '_1} \\
\frac{A-Q-\epsilon '_1}{4c '_1\epsilon '_1}, & \frac{-A-Q+\epsilon '_2}{4c '_2\epsilon '_2}, & \frac{A+Q+\epsilon '_2)}{4c '_3\epsilon '_2}, &\frac{-A+Q-\epsilon '_1}{4c '_4\epsilon '_1} \\
\frac{-A+Q+\epsilon '_1}{4c '_1\epsilon '_1}, & \frac{-A-Q+\epsilon '_2}{4c_2\epsilon '_2}, & \frac{A+Q+\epsilon '_2)}{4c '_3\epsilon '_2}, &\frac{A-Q+\epsilon '_1}{4c '_4\epsilon '_1}
\end{array}
\right), 
\label{X'-1}
\end{equation}
where 
\begin{eqnarray}
|c '_1|^2 &=& \frac{1}{2\left( \left(\frac{A-Q+\epsilon '_1}{B-P}\right)^2-1\right)} \label{c1'}, \\
|c '_2|^2 &=& \frac{1}{2\left( \left(\frac{A+Q+\epsilon '_2}{B+P}\right)^2-1\right)} \label{c2'}, \\
|c '_3|^2 &=& \frac{1}{2\left( \left(\frac{A+Q-\epsilon '_2}{B+P}\right)^2-1\right)} \label{c3'}, \\
|c '_4|^2 &=& \frac{1}{2\left( \left(\frac{A-Q-\epsilon '_1}{B-P}\right)^2-1\right)} \label{c4'}.
\end{eqnarray}
Then the dynamical structure factor is represented by
\begin{eqnarray}
S^{\alpha \beta}({\bm q}, \omega) &=& \sum_n S^{(n),\alpha \beta}(\bm q)\delta(\hbar \omega - \hbar \omega _{n}(\bm q)) \label{SQWBMGO}, \\
S^{(n)\alpha \beta}(\bm q) &=& S/4 \sum_{r,s=1}^{2}W_{r,\alpha}^{'(n)}(\bm q)W_{s,\beta}^{'(n)}(\bm q)^*, \\
W_{r,\alpha}^{'(n)}(\bm q) &=& V_{r, \alpha}^- X_{r,n}^{'-1}+V_{r, \alpha}^+X_{r+2,n}^{'-1}(\bm q), \\
V_{r,\alpha}^{\pm}&=&(U_r^{-1})_{\alpha, x} \pm i(U_r)^{-1}_{\alpha, y}. 
\end{eqnarray}
The inelastic neutron scattering cross section is expressed as follows: 
\begin{eqnarray}
\frac{d^2 \sigma}{d\Omega dE'} &=& \frac{(g _n r_0 /2)^2}{2\pi \hbar}\frac{k'}{k}|\frac{1}{2}gF(\bm q)|^2 \langle n_{\bm q}+1 \rangle \sum _{\alpha ,\beta = x,y,z}(\delta _{\alpha \beta} 
-\hat{\kappa _{\alpha}} \hat{\kappa _{\beta}}) S^{\alpha \beta}(\bm q, \omega), 
\label{crosssection} \\
\langle n_{\bm q}+1 \rangle &=& \frac{\exp (\hbar \omega _n ({\bm q})/k_BT)}{\exp (\hbar \omega_n ({\bm q})/k_BT) + 1}. 
\end{eqnarray}
Here, $g _n$ is $g$-factor of neutron, $r_0$ is the classical radius of electron, $k$ is the wave number of the incident neutron, 
$k'$ is the wave number of the scattered neutron, $g$ is Land{\'e}'s $g$-factor, $F(\bm q)$ is the magnetic form factor of the Mn$^{2+}$ ion~\cite{inttable}, 
and $\hat{\kappa _{\alpha}}$ is the $\alpha$ component of the unit scattering vector. 
$\langle n_{\bm q} \rangle$ is the temperature factor of the boson. 

The anisotropy energies at the magnetic $\Gamma$ point, $\bm q_{\Gamma} = (1, 0, 1/2)$, for the T$_1$ and T$_2$ modes are
\begin{eqnarray}
\Delta _1^{\rm SWT} &=& \hbar \omega_1(\bm q_{\Gamma}) = 4S^2\sqrt{2(2J_1+J_2+D/2)(8J_{\rm N}+4J_{\rm N2}\cos(4\kappa)-3a/2)} \label{Delta1SWT}, \\
\Delta _2^{\rm SWT} &=& \hbar \omega_2(\bm q_{\Gamma}) = 4S\sqrt{(2J_1+J_2)D} 
\end{eqnarray}
in the lowest order of $S$. 
In Eq.~(\ref{Delta1SWT}) $4\kappa = 90.4^{\circ}$, and $4J_{\rm N2}\cos(4\kappa)$ is negligible. 
As shown in Fig. 4(a) in the main article, $\Delta _2^{\rm SWT}$ is scaled by the sublattice moment, 
meaning that the single-ion anisotropy $D$ is temperature independent. 
This indicates that the coefficient $a$ and $b$ in the single-ion anisotropy with the 4th order term are 
also temperature independent, since 
the origins of the coefficients are commonly 
the products of matrix elements of the orbital angular momentum. 
$\Delta _1^{\rm SWT}$ in Fig. 4(b) in the main article is strongly dependent on the temperature. 
This suggests that the temperature independent component in $\Delta _1^{\rm SWT}$ is small, and 
$a$ is negligible. 
Then, $\Delta _1^{\rm SWT}$ is approximated as 
\begin{equation}
\Delta _1^{\rm SWT} \sim 16S^2\sqrt{J_{\rm N}(2J_1+J_2+D/2)}. \label{eq52}
\end{equation}
The coefficient $b$ does not appear in the leading term of the 
anisotropy energy of axial type $\Delta _1^{\rm SWT}$ in  Eq.~(\ref{Delta1SWT}). 
Thus, the single-ion anisotropy of the 4th order in Eq.~(\ref{H4}) can be neglected 
for the discussion of the magnetic anisotropy of axial type such as spin-flop field and the anisotropy gap 
in the INS spectrum. 

%%%%%%%%
%%%%%%%%
%%%%%%%%
%%%%%%%%
\section{Field derivative of magnetization}
%%%%%%%%
%%%%%%%%
%%%%%%%%
\begin{figure}[b]
\captionsetup{labelformat=empty,labelsep=none, justification=raggedright}
\includegraphics[width=8cm]{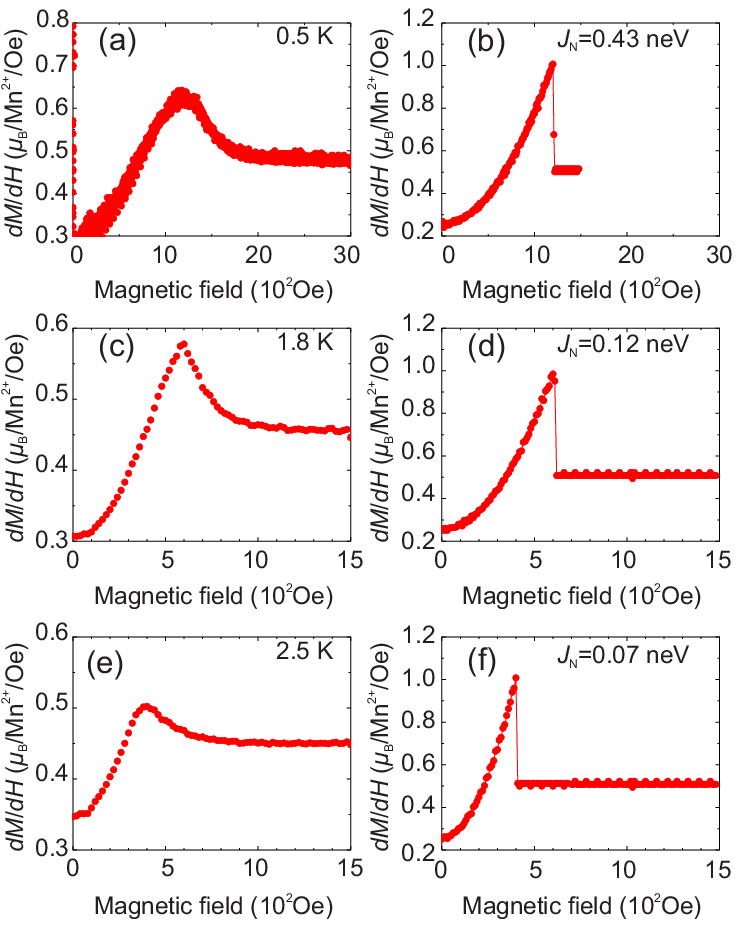}
\caption{{\bf Fig. S2} Field derivatives of magnetization ($dM$/$dH$) in the field applied along the [110] direction. 
(a) Measured $dM$/$dH$ at 0.5 K~\cite{Masuda10}. (b) Calculated $dM$/$dH$ with $J_{\rm N} = 0.43$ neV. 
(c) Measured $dM$/$dH$ at 1.8 K. (d) Calculated $dM$/$dH$ with $J_{\rm N} = 0.12$ neV. 
(e) Measured $dM$/$dH$ at 2.5 K. (f) Calculated $dM$/$dH$ with $J_{\rm N} = 0.07$ neV. 
}
\label{fig2}
\end{figure}
%%%%%%%%

\begin{figure}[t]
\captionsetup{labelformat=empty,labelsep=none, justification=raggedright}
\includegraphics[width=8cm]{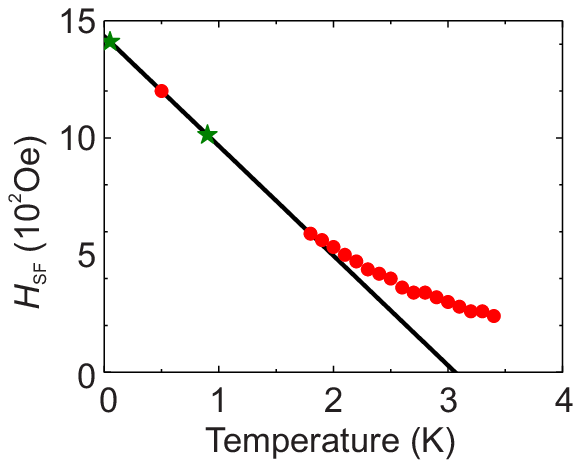}
\caption{{\bf Fig. S3} Estimate of the spin-flop field at 0.05 and 0.9 K. 
}
\label{fig3}
\end{figure}

In this section we estimate the spin-nematic interaction from the spin-flop field $H_{\rm SF}$. 
The field derivatives of magnetization $dM/dH$s in the field applied along the easy axis, $[110]$, 
at 0.5 K, 1.8 K, and 2.5 K are 
shown in Figs.~S\ref{fig2}(a), S\ref{fig2}(c) and S\ref{fig2}(e), respectively. 
We defined the peak of the $dM/dH$ as $H_{\rm SF}$, and they are shown in 
Fig. 2(c) in the main article. 
As for $T$ = 0.05 and 0.9~K, we estimated $H_{\rm SF}$ from the extrapolation and interpolation 
as shown in Fig.~S\ref{fig3}. 
They are used for the estimate of $\Delta _{1}^{\rm SWT}$ at the temperatures. 

%%%%%%%%
%%%%%%%%

%\begin{table}[p]
%\caption{\label{table1}Obtained parameters. {\bf ERROR BARS ARE REQUIRES, SOME DATA MISSING}}
%\begin{ruledtabular}
%\begin{tabular}{cccccc}
%$T$ (K)& 0.05 & 0.5 & 0.9 & 1.8 & 2.5 \\ 
%\colrule
%$g\langle S \rangle$ ($\mu_B$) & 4.68 & -- & 4.62 & 4.66 & 4.43 \\
%$J_{\rm{N}}^{\rm{eff}}$ (neV) & 0.64 & -- & 0.34 & 0.12 & 0.07 \\
%$\Delta ^{\rm HF}$ ($\mu$eV) & 31.3 & -- & 16.5 & 12.6 & 10.3 \\
%$H_{\rm SF}$ (T) & 0.141 & -- & 0.101 & 0.059 & 0.040 \\
%\end{tabular}
%\end{ruledtabular}
%\end{table}

%%%%%%%%
\begin{figure}[b]
\captionsetup{labelformat=empty,labelsep=none, justification=raggedright}
\includegraphics[width=8cm]{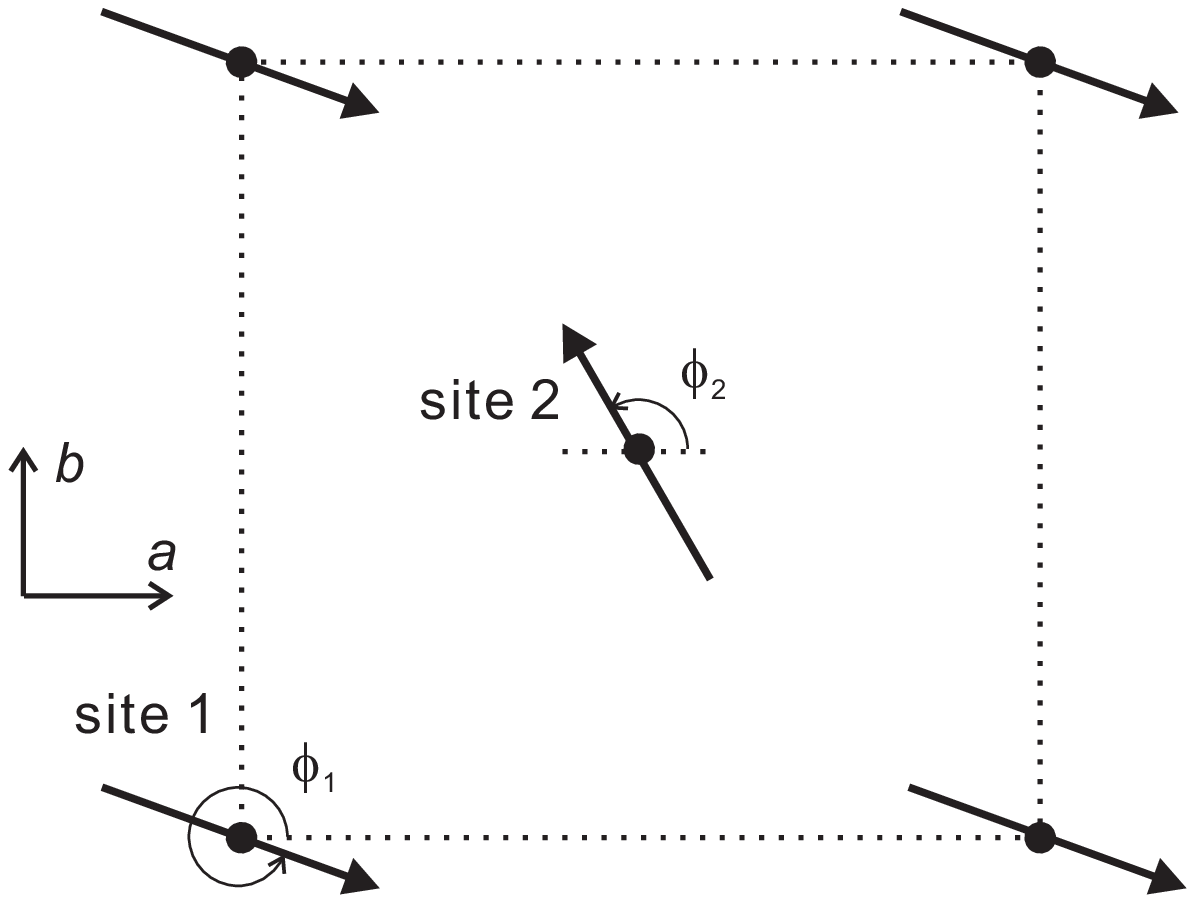}
\caption{{\bf Fig. S4} Spin configuration for the calculation of the energy. 
}
\label{fig4}
\end{figure}
%%%%%%%%

\begin{figure}[b]
\captionsetup{labelformat=empty,labelsep=none, justification=raggedright}
\includegraphics[width=8cm]{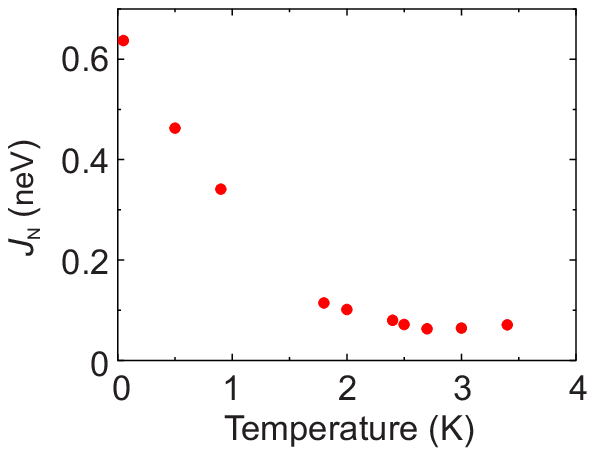}
\caption{{\bf Fig. S5} Temperature variation of evaluated spin-nematic interaction $J_{\rm N}$ from 
the analysis of $dM/dH$.  }
\label{fig2bb}
\end{figure}

\begin{figure}[t]
\captionsetup{labelformat=empty,labelsep=none, justification=raggedright}
\includegraphics[width=8cm]{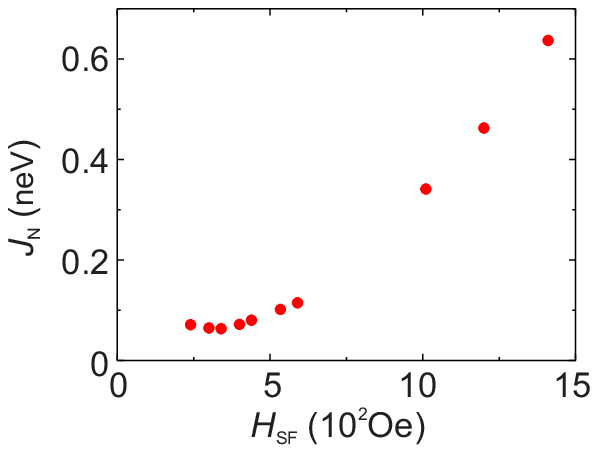}
\caption{{\bf Fig. S6} Relation between spin-flop field $H_{\rm SF}$ and spin-nematic interaction $J_{\rm N}$. 
}
\label{fig2aa}
\end{figure}

In order to calculate $dM/dH$, we numerically calculate the mean-field energy of the system, $E^{\rm MF}$, under the assumption that the spins are restricted in the 
crystallographic $ab$ plane as shown in Fig.~S\ref{fig4}. 
The spin structure is characterized by the angle between the $a$ axis and the spin direction, ${\phi}_r$. 
We consider the isotropic Heisenberg and spin-nematic terms as follows: 
\begin{eqnarray}
E^{\rm MF} &=& E^{\rm MF}_1 + E^{\rm MF}_2 + E^{\rm MF}_3, \label{Ec} \\
E^{\rm MF}_1 &=& \sum _{i,j}^{ab} J_1{\bm S}({\bm R}_i^{(1)}) \cdot {\bm S}({\bm R}_j^{(2)}) + \sum _{i,j}^{c} J_2{\bm S}({\bm R}_i^{(1)}) \cdot {\bm S}({\bm R}_j^{(2)}) \\
E^{\rm MF}_2 &=& J_{\rm N} \sum_{i,j}^{ab} O_{XY}({\bm R}_i^{(1)})O_{XY}({\bm R}_j^{(2)}) \\ \nonumber
&& O_{XY}({\bm R}_i^{(r)}) = \cos (2\kappa_i^{(r)})O_{xy}({\bm R}_i^{(r)}) - \sin (2\kappa_i^{(r)})O_2^2({\bm R}_i^{(r)}) \\ \nonumber
&& O_{xy}({\bm R}_i^{(r)}) = S^x({\bm R}_i^{(r)})S^y({\bm R}_i^{(r)}) + S^y({\bm R}_i^{(r)})S^x({\bm R}_i^{(r)}) \\ \nonumber
&& O_2^2 ({\bm R}_i^{(r)}) = (S^x({\bm R}_i^{(r)}))^2 - (S^y({\bm R}_i^{(r)}))^2 \\ 
E^{\rm MF}_3 &=& -g{\mu}_B\sum_{i,k} {\bm S}({\bm R}_i^{(r)}) \cdot {\bm H} \\ 
{\bm S}({\bm R}_i^{(r)}) &=& S(\cos {\phi}_r, \sin {\phi}_r, 0). 
\end{eqnarray}
From the discussion in the Section I, we neglect $J_{\rm N2}$, $a$ and $b$ in the original Hamiltonian in 
Eq.~(\ref{hamiltonian}). 
We use the magnitudes of the spin moments at 0.5, 1.8 and 2.5 K 
which are obtained from the present experiment. % as summarized in Table I in the main text. 
We use $J_1$ = 27.8 ${\mu}$eV, $J_2 = 1~{\mu}$eV and $g$ = 2. 
$J_{\rm N}$ is a free parameter. 
We numerically calculate $E^{\rm MF}$ of the spin configuration with a $J_{\rm N}$ 
for a set of random parameters ${\phi}_r$, and 
we find the spin structure having the minimum energy. 
$dM/dH$s are calculated for the spin structures with sequential $J_{\rm N}$s with a step of 0.001 neV. 
The best $dM/dH$s that reproduce the spin-flop fields at $J_{\rm N}$s are obtained at the temperatures, 
and they are shown in Figs.~S\ref{fig2}(b), S\ref{fig2}(d), and S\ref{fig2}(f). 
$J_{\rm N}$ is strongly dependent on temperature as shown in Fig.~S\ref{fig2bb}. 
The relation between $H_{\rm SF}$ and $J_{\rm N}$ is shown in Fig.~S\ref{fig2aa}. 
The positive correlation is found, and $H_{\rm SF}$ is a good index for the energy of the axial magnetic anisotropy.

%%%%%%%%
%%%%%%%%

%%%%%%%%
%%%%%%%%

\section{Analysis of inelastic neutron scattering spectra}
Inelastic neutron scattering spectra in Fig.~3(f) in the main article is analyzed by Eqs.~(\ref{SQWBMGO}) and (\ref{crosssection}). 
The broadening of the peak is due to both instrumental resolution and finite lifetime of magnon. 
Here we do not distinguish them, and we introduce Gaussian function. 
The $\delta$ function in Eq.~(\ref{SQWBMGO}) is, then, replaced by 
\begin{equation}
\delta(\hbar \omega - \hbar \omega_n(\bm q)) \rightarrow \frac{1}{\sqrt{\pi} \Gamma}
\exp\left(-\left(\frac{\hbar \omega - \hbar \omega_n(\bm q)}{\Gamma}\right)^2\right), 
\label{width}
\end{equation}
where $\Gamma$ is the broadening factor. 
As derived in Eqs.~(\ref{DeltaHF}) and (\ref{eq52}), $J_{\rm N}$, $J_1$, $J_2$, $D$ and $\Delta^{\rm HF}$ 
contribute to the energy gap. 
%Both ${\cal H}_3$ in Eq.~(4) and ${\cal H}_4$ in Eq.~(\ref{H4}) contribute to the anisotropy gap . 
%and they cannot be distinguished in our analysis. 
$N_2^{\rm eff}$, $a$, and $b$ are irrelevant, and we fix $N_2^{\rm eff} = a = b = 0$. 
We use $J_1 = 27.8~\mu{\rm eV}$ and $J_2 = 1~\mu{\rm eV}$ from previous study~\cite{Masuda10}. 
As for $J_{\rm N}$ in Eq.~(\ref{H2}) we use the values estimated from the spin-flop field. 
The free parameters are $D$ in Eq.~(3), $\Delta ^{\rm HF}$ in 
Eq.~(\ref{DeltaHF}), and $\Gamma$ in Eq.~(\ref{width}). 
Fits to the data are shown in Figs.~3(f) in the main article, and the obtained parameters 
are summarized in Table~S\ref{tab2}. 

\begin{table}[t]
\captionsetup{labelformat=empty,labelsep=none, justification=raggedright}
\caption{{\bf Table SI} $D$ and $\Gamma$ estimated from INS spectra. }
\begin{ruledtabular}
\begin{tabular}{ccccc}
$T$ (K)& 0.05 & 0.9 & 1.8 & 2.5 \\ 
\colrule
$D$ ($\mu$eV) & 2.4 & 2.4 & 2.2 & 2.1 \\
%$\Delta ^{\rm HF}$ ($\mu$eV) & 29.35 & -- & 14.05 & 7.99 & 0.083 \\
$\Gamma$ ($\mu$eV) & 6.7 & 10 & 16 & 19
\label{tab2}
\end{tabular}
\end{ruledtabular}
\end{table}

%%%%%%%%
%\begin{figure}[p]
%\captionsetup{labelformat=empty,labelsep=none, justification=raggedright}
%\includegraphics[width=18cm]{sfig5.eps}
%\caption{{\bf Fig. S7} Energy spectra sliced at $\bm q$ = (1, 0, 0.5) at $T$ = 0.05 K (a), 0.9 K (b), 1.8 K (c), 
%and 2.5 K (d). 
%Black open circles are the data and blue curves are the calculation. }
%\label{fig5}
%\end{figure}
%%%%%%%%

\section{Temperature dependence of $\Delta ^{\rm HF}$}
\begin{figure}[b]
\captionsetup{labelformat=empty,labelsep=none, justification=raggedright}
\includegraphics[width=8cm]{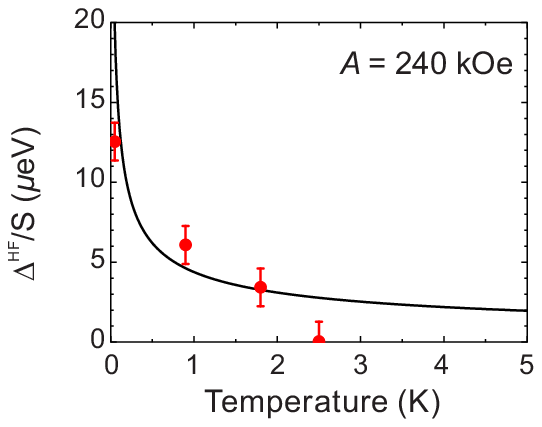}
\caption{{\bf Fig. S7} Temperature dependence of $\Delta ^{\rm HF}/\langle S(T) \rangle$ estimated from INS spectra. 
}
\label{fig6}
\end{figure}

The energy gap induced by hyperfine coupling between nuclear spin and electron spin is expressed 
as follows: 
\begin{equation}
\Delta^{\rm{HF}}=\sqrt{ ( \chi_n(T) / \chi_c(T) )} Ag {\mu}_B \langle S(T) \rangle. 
\end{equation}
Here $\chi_c(T)$ is the susceptibility of electron spin in the 
field applied along the $c$ axis, and $A$ is the hyperfine coupling constant. 
The $c$ axis is perpendicular to the magnetic easy axis, and $\chi_c(T)$ is almost independent on the temperature. 
Then $\chi_c = $ 0.25 emu/mole~\cite{Masuda10} is used. 
$\chi_n(T)$ is the susceptibility of nuclear spin, yielding to Curie law
\begin{eqnarray}
\chi_n(T)&=&\frac{C}{T} \\
C&=&\frac{Ng_N^2\mu _N^2 I(I+1)}{3k_B}. 
\end{eqnarray}
$(\Delta ^{\rm HF}/\langle S(T) \rangle)^2$ is proportional to $\chi_n(T)$, and it is shown in Fig.~S\ref{fig6}. 
It is estimated that $A = 240$ kOe. 
In Mn materials, $A$ is reported to be 259 kOe~\cite{JETPL.64.473,Clogston}, and our value is close to it. 

%%%%%%%%
%%%%%%%%